\documentclass[aps,prd,superscriptaddress,nofootinbib,groupedaddress,showkeys]{revtex4}
\usepackage{graphicx}  
\usepackage{dcolumn}   
\usepackage{bm}        
\usepackage{amssymb}   
\usepackage{amsmath}
\usepackage{xcolor}
\usepackage{amsthm}
\newtheorem{theorem}{Theorem}

\newcommand{\be}{\begin{equation}}\newcommand{\ee}{\end{equation}}
\newcommand{\bea}{\begin{eqnarray}}\newcommand{\eea}{\end{eqnarray}}
\newcommand{\brr}{\begin{array}}\newcommand{\err}{\end{array}}
\newcommand{\bit}{\begin{itemize}}\newcommand{\eit}{\end{itemize}}
\newcommand{\ben}{\begin{enumerate}}\newcommand{\een}{\end{enumerate}}

\newcommand{\bbm}{\begin{bmatrix}}\newcommand{\ebm}{\end{bmatrix}}
\newcommand{\ba}{\begin{array}}
\newcommand{\ea}{\end{array}}
\newcommand{\G}{\textbf}

\newtheorem{mydef}{Definition}
\newtheorem{Lemma}{Lemma}
\newcommand{\bd}{\begin{mydef}} \newcommand{\ed}{\end{mydef}}
\newcommand{\bthe}{\begin{theorem}} \newcommand{\ethe}{\end{theorem}}
\newcommand{\ble}{\begin{Lemma}} \newcommand{\ele}{\end{Lemma}}

\newcommand{\dr}{\mathrm{d}}

\definecolor{darkred}{rgb}{.8,0,0}

\definecolor{darkblue}{rgb}{0,0,.7}

\def\ha{\frac{1}{2}}

\def\intx{\int \!\!\mathrm{d}^3 {\G x}}

\def\ph{\varphi}
\def\lan{\langle}
\def\lf{\left}

\def\non{\nonumber}\def\pa{\partial}\def\ran{\rangle}

\def\ri{\right}
\def\al{\alpha}\def\bt{\beta}
\def\de{\delta}\def\De{\Delta}
\def\te{\theta}
\def\la{\lambda}\def\La{\Lambda}\def\si{\sigma}
\def\om{\omega}\def\Om{\Omega}

\def\1{{_{1}}}\def\2{{_{2}}}

\newcommand{\ide}{1\hspace{-1mm}{\rm I}}

\def\noHe0{:\;\!\!\;\!\!:H_e(0):\;\!\!\;\!\!:}
\def\noHm0{:\;\!\!\;\!\!:H_\mu(0):\;\!\!\;\!\!:}

\def\lan{\langle}
\def\lf{\left}

\def\non{\nonumber}
\def\pa{\partial}\def\ran{\rangle}

\def\ri{\right}

\def\al{\alpha}\def\bt{\beta}
\def\de{\delta}\def\De{\Delta}
\def\te{\theta}
\def\la{\lambda}
\def\La{\Lambda}\def\si{\sigma}
\def\om{\omega}\def\Om{\Omega}

\def\1{{_{1}}}\def\2{{_{2}}}

\begin{document}

\preprint[\leftline{KCL-PH-TH/2020-{\bf 03}}

\title{Some non-trivial aspects of Poincar\'e  and $CPT$ invariance of flavor vacuum}
\author{M.~Blasone}
\email{blasone@sa.infn.it}
\affiliation{Dipartimento di Fisica, Universit\`a di Salerno, Via Giovanni Paolo II, 132 84084 Fisciano, Italy \& INFN Sezione di Napoli, Gruppo collegato di Salerno, Italy}
\author{P.~Jizba}
\email{p.jizba@fjfi.cvut.cz}
\affiliation{FNSPE, Czech Technical
University in Prague, B\v{r}ehov\'{a} 7, 115 19 Praha 1, Czech Republic\\}
\affiliation{ITP, Freie Universit\"{a}t Berlin, Arnimallee 14,
D-14195 Berlin, Germany}
\author{N.E.~Mavromatos}
\email{Nikolaos.Mavromatos@kcl.ac.uk}

\affiliation{Theoretical Particle Physics and Cosmology Group, Department of Physics, King's College London, Strand WC2R 2LS, UK}

\author{L.~Smaldone}
\email{smaldone@ipnp.mff.cuni.cz}
\affiliation{Faculty  of  Mathematics  and  Physics, Charles  University, V  Hole\v{s}ovi\v{c}k\'{a}ch  2, 18000  Praha  8,  Czech  Republic}

\begin{abstract}
We study the explicit form of Poincar\'e and discrete transformations of flavor states in
a two-flavor scalar model, which represents the simplest example of the field mixing. Because of the particular form of the flavor vacuum condensate, we find that
the aforementioned symmetries are spontaneously broken. The ensuing vacuum stability group is identified with the Euclidean group $E(3)$. With the help of Fabri--Picasso theorem, we show that flavor vacua with different time labels and in different Lorentz frames are unitarily inequivalent to each other and they constitute a manifold of zero-flavor-charge states. Despite the spontaneous breakdown of Poincar\'e and CPT symmetries that characterises such vacua, we 
provide arguments on the absence of Goldstone Bosons. We also prove that the phenomenologically relevant oscillation formula is invariant under these transformations.
\end{abstract}

\maketitle
\section{Introduction}
The fundamental particles are usually classified, following Bargmann and Wigner~\cite{BarWig}, in terms of unitary irreducible representations of Poincar\'e group~\cite{StWi,BogLog}. According to this classification, particles and ensuing vacuum states are characterized by their mass $m$ and spin $s$ (or helicity, in the case of massless particles). In the case of particles without a sharp value of mass (e.g. unstable particles), such a classification can be regarded, at best, as an approximation~\cite{BogLog}. In such cases the concept of sharp mass is substituted with a finite mass-width  distribution. Ensuing variance is proportional to the inverse of particle half-life due to time-energy uncertainty relation~\cite{Perkins:2000,uns}. This picture can also be explained in terms of a non-trivial vacuum structure possessed by such systems~\cite{DeFVit}.

It was recently pointed out (see Ref.~\cite{BJSun}) that flavor neutrino states share a common feature with unstable particles, in the sense that only their energy (mass) distribution has a physical meaning and the width of this distribution is related to the inverse of the oscillation length which can be again deduced from time-energy uncertainty relation~\cite{BJSun,Bil}. Furthermore, the latter result was recently generalized, in a quantum mechanical context, to stationary curved spacetimes \cite{Blasone:2019jtj}.

It is worthwhile therefore to clarify the relation between flavor states and unitary representations of Poincar\'e group. This point was first tackled in Ref.~\cite{Lobanov}, where it was proposed to extend the Poincar\'e group so as to include an internal $SU(3)$ flavor symmetry in the Standard Model. Because of \emph{Coleman--Mandula theorem}~\cite{ColMan}, the extended group can only be $T^{3,1} \rtimes O(3,1) \times SU(3)$.

In parallel, Lorentz invariance properties of neutrino oscillation formula were studied in a number of papers, e.g.~\cite{Giunti:2003ku,maguejo,BLDiMLa}.  In Ref.~\cite{Giunti:2003ku}, the invariance of the standard oscillation formula~\cite{Pontecorvo}, was explicitly proved but only in the ultrarelativistic case. However, it can be argued~\cite{Giuntibook,NeutPheno,MixingReview,BlaVit95,qftmixing} that such a formula should be regarded as a Quantum-Mechanical (QM) approximation of a more fundamental Quantum Field Theoretical (QFT) relation.  In particular QM behavior cannot grasp the non-trivial vacuum structure whose observable effects are more important at lower energies. This point was deeply analyzed within the framework of the so-called {\it flavor vacuum} quantization of field theories with mixing~\cite{BlaVit95}, which entails the important features that the Fock space of states with definite flavor is unitarily inequivalent to the Fock space of states with definite mass, and
 that  the {\it flavor vacuum} is structurally similar to that of a BCS condensate. Within this scenario, in Ref.~\cite{BLDiMLa} a preliminary study of the QFT oscillation formula in different Lorentz frames was undertaken and Lorentz violating effects were identified~\cite{BlaVit95,qftmixing}. In passing we remark that low-energy neutrino phenomenology is becoming increasingly important in understanding cosmic
neutrino background (CNB) and its potential cosmological implications~\cite{Follin:2015,Ringwald:2004}.

Within this QFT treatment of neutrino oscillations, associated deformations of the Lorentz energy dispersion relations were studied  in Ref.~\cite{maguejo}. Such a modification of the dispersion dispersion relations for the flavor states can be understood as an ``environmental'' effect of quantum-gravitational degrees of freedom
in a concrete model of quantum gravity within the framework of string/brane theory, the so-called
 \emph{D}-foam model~\cite{Ellis:2004ay}. In this context, the scattering between open strings, representing flavored matter, and $D0$-branes, which are viewed as Poincar\'e-symmetry-breaking point-like space-time defects, of quantum-gravitational stringy nature, is modelled by an effective theory, which entails the dynamical generation of mixing via flavor vacuum condensates~\cite{mavroman,DynMix}, in agreement with the generic feature of the flavor vacuum~\cite{BlaVit95}, mentioned previously.
 Such  a vacuum structure  can also be obtained  via algebraic, i.e., non perturbative methods, in the case of two~\cite{ccpaper} and three flavor~\cite{ccpaper3} models with $SU(n)_L \times SU(n)_R \times U(1)_V$ chiral flavor symmetry.

In this connection we can also point out that Lorentz violating effects implied by a fundamental string theoretical dynamics characterize also the Standard Model extension (SME) proposed by Colladay and Kostelecky~\cite{Colladay:1996iz}. In the SME, $CPT$ and Lorentz violating terms are explicitly added to the Standard Model (SM) Lagrangian.
At this stage, we should also like to recall the work of \cite{cptgreen}, according to which a violation of CPT {\it necessarily} implies the violation of Lorentz invariance.
Within such a framework the neutrino oscillations were studied in~\cite{Kostelecky2004} and modified dispersion relations connected with an underlying Planck scale physics were found. Following these developments, many authors dedicated their efforts to an understanding of both the theoretical and the phenomenological implications of SME or SME-like scenarios (see e.g.~\cite{ColGla,Katori:2012pe,Antonelli:2018fbv}). It can  also be argued~\cite{Lambiase:2017adh} that bounds on the parameters of SME can be fixed through generalized uncertainty principle~\cite{GUP}.

The aim of this paper is to study Poincar\'e and discrete symmetries in a simple toy model that describes oscillation of a two flavor ($A,B$) scalar field doublet with mixing~\cite{Binger:1999nj,bosons}. In this context we propose yet another solution to the apparent incompatibility of Poincar\'e symmetry on flavor states, namely that the Poincar\'e symmetry is spontaneously broken  on \emph{flavor vacuum}~\cite{BJSun,BlaVit95,qftmixing,Binger:1999nj,bosons}. So, in particular, the Lagrangian symmetry does not leave vacuum invariant and the residual symmetry is found to be $E(3)$. This spontaneous symmetry breakdown (SSB) is caused by the complicated condensate structure of the flavor vacuum. Here we do not specify the origin of this condensate, which can be motivated by physics beyond SM as is done, e.g., in Refs.~\cite{mavroman,DynMix,ccpaper}. This would, in turn, indicate the necessity for a dynamical origin of mixing. The action of the broken charges as symmetry generators on the vacuum, defines a linear manifold of flavor-degenerate states, which represent the~\emph{flavor vacuum manifold}. All points on such a vacuum manifold represent unitarily inequivalent Fock spaces.
With the same reasoning we prove that $CPT$ symmetry is also spontaneously broken on the flavor vacuum, with the residual symmetry being $CP$. In view of the theorem in~\cite{cptgreen}, then, the breaking of Lorentz symmetry by the flavor vacuum can be attributed to the (spontaneous) breaking of $CPT$ symmetry in this approach.

As a main result, we prove, quite surprisingly, that such a violation does not affect the phenomenologically relevant flavor oscillation formula, which is demonstrated to be Poincar\'e  invariant. In fact, here we employ a wave-packet approach for neutrino oscillations developed in Ref.~\cite{Blasone:2002wp}, which permits to treat this issue in a manifestly covariant way. The same result can be derived for continuous time-translations, $T$ and $CPT$ transformations.

The present paper is organized as it follows: in Section~\ref{pomix} we discuss the incompatibility of irreducible representations of the Poincar\'e group on flavor states. In Section~\ref{Sec2} the canonical quantization of flavor (scalar) fields is reviewed~\cite{Binger:1999nj,bosons} and we set up convention employed in the rest of the paper. Here, unlike in Refs.~\cite{Binger:1999nj,bosons}, we use the invariant form of canonical commutation relations, which makes more evident eventual Lorentz violations. In Section~\ref{PoiSSB} Poincar\'e group generators are explicitly constructed, in the flavor representations, and SSB of time-translations and Lorentz boosts is shown. Then, in Section \ref{Sec4}, the same  procedure is repeated for the case of discrete symmetries, showing that $CPT$ is broken on the flavor vacuum. Finally, in Section~\ref{conclusions}, conclusions and future perspectives are presented. For reader's convenience we include two appendices that complement more technical aspects from the main text.

\section{Poincar\'e group representations and field mixing}\label{pomix}

In this section we briefly discuss the problem of constructing flavor states in connection with unitarily irreducible representations of Poincar\'e group. By using the commutation relations \eqref{poal1}-\eqref{poal3} one can verify that Poincar\'e group has two Casimir invariants~\cite{StWi,BogLog}:
\be
M^2 \ \equiv \ P_\si P^{\si} \, \qquad W^2 \ = \ W_\si W^\si \, ,
\ee
where
\be
W_\si \ = \ -\ha \, \varepsilon_{\mu \nu \rho \si} \, J^{\mu \nu} \, P^\rho \, ,
\ee
is the \emph{Pauli--Lubansky} operator.

After Bargmann and Wigner \cite{BarWig}, particle states are usually assumed to belong to the unitary irreducible representations of the Poincar\'e group \cite{StWi,BogLog}. It follows that the two quadratic Casimir invariants act as a multiple of the identity operator, on these states:
\be
M^2 \, |k^2,s,\si\ran \ = \  m^2_\si \, |k^2,s,\si\ran \, , \qquad W^2 \, |k^2,s,\si\ran \ = \  \,- m^2_\si \, s (s+1)  \, |k^2,s,\si\ran \, ,
\ee
where $\si$ indicates some quantum number (e.g. flavor), $m_\si$ is the particle mass and $s$ is its spin\footnote{Here, for simplicity, we assume the same spin for each $\si$.}.

Let us now consider the Lagrange density
\be \label{lagden}
\mathcal{L}(x)\ = \ \pa_\mu \boldsymbol{\ph}^\dagger_f(x) \pa^\mu \boldsymbol{\ph}_f(x) \ -\ \boldsymbol{\ph}^\dagger_f(x) \, M^2 \, \boldsymbol{\ph}_f(x) \, ,
\ee
where
\be \label{lagrflavfields}
\boldsymbol{\ph}_f(x)\ =\ \bbm \ph_A(x)  \\ \ph_B(x)  \ebm \, , \qquad M^2\ = \ \bbm m^2_A & m^2_{A \, B} \\ m^2_{A \, B} & m^2_B \ebm \, ,
\ee
which describes the dynamics of two coupled (mixed) scalar fields that we will call {\em flavor fields}, in a close analogy with the terminology used in quark and neutrino physics. A pressing problem in the study of fundamental aspects of flavor physics is the correct definition of flavor states \cite{NeutPheno,MixingReview,BlaVit95,qftmixing}. However, it is clear that these cannot be taken as elements of irreducible representations of the Poincar\'e group. This was already noticed, e.g., in Ref. \cite{Lobanov}. The argument for this is very simple: if this were not true we should have\footnote{Here we do not consider the Pauli--Lubansky operator because we limit to the case of scalar (spinless) fields.}:
\be
M^2 \, |k_\si,\si\ran \ = \  m^2_\si \, |k_\si,\si\ran \, , \qquad \si \ = \ A,B \, .
\label{5bc}
\ee
which is clearly false, because flavor states do not have a definite mass\footnote{Strictly speaking, in QM one can construct an operator of the form (\ref{5bc}) but such operator cannot be interpreted as a mass operator. In QFT this is impossible due to unitary inequivalence of flavor and mass representation.}.

It thus seems that Poincar\'e symmetry is not compatible with flavor mixing. One possibility would be to extend the Poincar\'e group. For instance, in Ref.~\cite{Lobanov} it was proposed to consider $T^{3,1} \rtimes \, O(3,1) \times SU(n)$, where $n$ is the number of flavors involved. In sections to follow we propose and discuss yet another possibility, namely we will quantize flavor fields directly in the flavor space, where the vacuum is manifestly Poincar\'e non-invariant and show that the Poincar\'e symmetry is spontaneously broken in the symmetry breaking scheme
\be
T^{3,1} \rtimes \, O(3,1)  \ \rightarrow \ E(3) \, .
\ee
Here $E(3)$ denotes the three dimensional Euclidean group.

The present analysis does not investigate the actual mechanism that is responsible for this SSB. A simple dynamical model where such a SSB can naturally be encountered is considered in Ref.~\cite{ccpaper}. There it is shown that a \emph{necessary} condition for dynamical generation of fermion mixing, in models characterized by \emph{chiral flavor symmetry}, is the vacuum condensation of fermion-antifermion pairs, which mix particles with different masses and so, \emph{dynamical mixing generation requires a mixing at level of vacuum}. In that context, the Lorentz symmetry is spontaneously broken by the presence of such exotic condensates, via the SSB scheme:
\be
T^{3,1} \rtimes \, O(3,1) \times SU(2)_L \times SU(2)_R \times U(1)_V \rightarrow U(1)_V \times E(3)\, ,
\ee
where $L$ and $R$ indicate the left and right components of the chiral group, respectively, and $V$ is the vector group. The global $U(1)_V$ invariance is related to the conservation of total flavor charge. Here we believe that it is quite feasible that a similar mechanism drives the SSB of Poincar\'e symmetry also in the bosonic case.

\section{Flavor fields quantization}\label{Sec2}

Let us now consider a simple scalar model for flavor oscillations described by the Lagrange density~\eqref{lagden}, which can be diagonalized through the following transformation:
\bea \label{bosmix}
\boldsymbol{\ph}_f(x)\ =\ U \,  \boldsymbol{\ph}_m(x) \, , \qquad U=\bbm \cos \te & \sin \te \\ -\sin \te & \cos \te \ebm  \, ,
\eea
where $\tan 2 \theta \ = \ 2 \, m^2_{AB}/\lf(m^2_B-m^2_A\ri)$. After this transformation, $\mathcal{L}$ becomes
\be
\mathcal{L}(x)\ = \ \pa_\mu \boldsymbol{\ph}^\dagger_m(x) \,  \pa^\mu \boldsymbol{\ph}_m (x)\ -\ \boldsymbol{\ph}^\dagger_m(x) \, M^2_d \,  \boldsymbol{\ph}_m (x) \, ,
\label{II.4.a}
\ee
where
\be \label{lagrmassfields}
\boldsymbol{\ph}_m(x)\ =\ \bbm \ph_1(x) \\ \ph_2(x)  \ebm \, , \qquad M^2_d\ =\ \bbm m^2_1 & 0 \\ 0 & m^2_2 \ebm \, .
\ee
The Lagrange density (\ref{II.4.a}) describes two free scalar fields with definite particle masses $m_1$ and $m_2$. They can be thus expanded as:
\bea
\ph_j(x)& = & \int \frac{\dr^3 \G k}{2\om_{\G k,j}(2 \pi)^3} \, \lf[a_{\G k,j}\, e^{-i\om_{\G k,j} t}\ + \ b^{\dagger}_{-\G k,j} \, e^{i\om_{\G k,j} t}\ri] \, e^{i\G k \cdot \G x} \, , \quad j=1,2\, ,
\eea
where the annihilation and creation operators satisfy the following commutation relations:
\be \label{liccr}
\lf[a_{\G k,i} \, , \, a^\dag_{\G p,j} \ri] \ = \ \lf[b_{\G k,i} \, , \, b^\dag_{\G p,j} \ri] \ = \ 2 \om_{\G k,i} \, (2 \pi)^3 \, \de(\G k-\G p) \,  \de_{ij} \, ,
\ee
and annihilate the \emph{mass vacuum}:
\be
a_{\G k,j}|0\ran_{1,2} \ = \ b_{\G k,j}|0\ran_{1,2} \ = \ 0 \, ,
\ee
i.e., the ground state of the system. Note that, in contrast to Refs.~\cite{Binger:1999nj,bosons} we use the Lorentz invariant commutation relations~\eqref{liccr}.
We now expand flavor fields in a similar way:
\bea \label{expph}
\ph_\si(x)\ = \ \int \frac{\dr^3 \G k}{2\om_{\G k,\si}(2 \pi)^3} \, \lf[a_{\G k,\si}(t)\, e^{-i\om_{\G k,\si} t}\ + \ b^{\dagger}_{-\G k,\si}(t) \, e^{i\om_{\G k,\si} t}\ri] \, e^{i\G k \cdot \G x} \, , \quad \si=A,B\, ,
\eea
with $\om_{\G k,\si}= \sqrt{|\G k|^2+\mu_\si^2}$ and $\mu_\si$ are mass parameters which have to be specified. From the mixing transformation~\eqref{bosmix} it follows that\footnote{Here the time dependence of creation and annihilation operators indicates that flavor fields are interacting fields. Actually, this interacting model can be solved exactly, without perturbation expansion.}:
\be
a_{\G k,A}(t) \ = \ \intx \, e^{i \lf(\om_{\G k,A} t-\G k \cdot \G x\ri)} \, i \stackrel{\leftrightarrow}{\pa_0} \lf(\cos \theta \, \ph_1(x) \, + \, \sin \theta \, \ph_2(x)\ri) \, ,
\ee
and similarly for other operators. Explicitly, we find that
\bea
  \left[ \begin{tabular}{c} $a_{\G k,A}$ \\ $b_{-\G k,A}^{\dagger}$
\\$a_{\G k,B}$ \\ $b_{-\G k,B}^{\dagger}$ \end{tabular} \right]
= \left[\begin{array}{cccc}
c_\theta\, \rho^{\G k}_{A 1}& c_\theta \, \lambda^{\G k}_{A 1} &
s_\theta \,\rho^{\G k}_{A 2}  &
s_\theta \,\lambda^{\G k}_{A 2}
\\  c_\theta \,\lambda^{\G k}_{A 1} & c_\theta\, \rho^{\G k}_{A 1} &  s_\theta
\,\lambda^{\G k}_{A 2} & s_\theta \,\rho^{\G k}_{A 2}
\\ - s_\theta \,\rho^{\G k}_{B 1} &-s_\theta \,\lambda^{\G k}_{B 1}& c_\theta
\,\rho^{\G k}_{B 2}
&  c_\theta \,\lambda^{\G k}_{B 2} \\  - s_\theta \,\lambda^{\G k}_{B 1} &-
s_\theta\,
\rho^{\G k}_{B 1} &  c_\theta\, \lambda^{\G k}_{B 2}& c_\theta\, \rho^{\G k}_{B 2}
\end{array}\right]
  \left[ \begin{tabular}{c} $a_{\G k,1}$ \\ $b_{-\G k,1}^{\dagger}$ \\  $a_{\G k,2}$ \\ $b_{-\G k,2}^{\dagger} $\end{tabular} \right] \, ,
 \label{4x4Bog}
\eea
where
$c_\theta\equiv \cos\theta$, $s_\theta\equiv \sin\theta$, and
\be
\rho^{\G k}_{\si j} \ = \ |\rho^{\G k}_{\si j}|e^{i (\om_{\G k,\si}-\om_{\G k,j})t} \, , \qquad \la^{\G k}_{\si j} \ = \ |\la^{\G k}_{\si j}|e^{i (\om_{\G k,\si}+\om_{\G k,j})t} \,  , \quad (\si,j)=(A,1),(B,2)\, ,
\ee
where
\be
|\rho^{\G k}_{\si j}| \ = \ \ha \lf(\frac{\om_{\G k,\si}}{\om_{\G k,j}}+1\ri) \, , \qquad |\la^{\G k}_{\si j}| \ = \ \ha \lf(\frac{\om_{\G k,\si}}{\om_{\G k,j}}-1\ri) \, .
\ee
Note that~\eqref{4x4Bog} represents a canonical transformation because
\be \label{flavccr}
\lf[a_{\G k,\si}(t) \, , \, a^\dag_{\G p,\rho}(t) \ri] \ = \ \lf[b_{\G k,\si}(t) \, , \, b^\dag_{\G p,\rho} (t)\ri] \ = \ 2 \om_{\G k,\si} \, (2 \pi)^3 \, \de(\G k-\G p) \,  \de_{\si \rho} \, .
\ee
For future convenience, we write explicitly the inverse transformation as:
\bea
 \boldsymbol{a}_{\G k,j} \ = \  \sum_{\si=A,B} \, J^{\G k}_{j,\si}(t) \, \boldsymbol{a}_{\G k,\si}(t) \, ,
 \label{4x4Boginv}
\eea
where $\boldsymbol{a}_{\G k,j}= \lf[a_{\G k,j} \ b_{-\G k,j}^{\dagger}\ri]^T$, $\boldsymbol{a}_{\G k,\si}= \lf[a_{\G k,\si} \ b_{-\G k,\si}^{\dagger}\ri]^T$, and the matrix $J^\G k$ has the form
\be
J^{\G k}(t) \ \equiv \ \left[\begin{array}{cccc}
c_\theta\, \rho^{\G k}_{1A}& c_\theta \, \lambda^{\G k}_{1A} &
-s_\theta \,\rho^{\G k}_{1 B}  &
-s_\theta \,\lambda^{\G k}_{1 B}
\\  c_\theta \,\lambda^{\G k}_{1A} & c_\theta\, \rho^{\G k}_{1A} &  -s_\theta
\,\lambda^{\G k}_{1 B} & -s_\theta \,\rho^{\G k}_{1 B}
\\ s_\theta \,\rho^{\G k}_{2 A} &s_\theta \,\lambda^{\G k}_{2 A}& c_\theta
\,\rho^{\G k}_{2 B}
&  c_\theta \,\lambda^{\G k}_{2 B} \\  s_\theta \,\lambda^{\G k}_{2 A} &
s_\theta\,
\rho^{\G k}_{2 A} &  c_\theta\, \lambda^{\G k}_{2 B}& c_\theta\, \rho^{\G k}_{2 B}
\end{array}\right] \ = \ \bbm J^\G k_{1 A}(t) & J^\G k_{1 B}(t) \\[1mm] J^\G k_{2 A}(t) & J^\G k_{2 B}(t)\ebm\, ,
\ee
and $J^{\G k}_{j \si}$ are $2 \times 2$ symmetric matrices.

Let us notice that we have not specified the mass parameters $\mu_\si$. The situation here is similar to the one encountered in QFT in curved spacetime~\cite{Curved} where one has an infinite set of creation and annihilation operators related by a Bogoliubov transformation. In Ref.~\cite{Casimiro} it was shown that different choices of $\mu_\si$ affect the strength of the Casimir force between two plates. Typical choices studied in literature~\cite{Blasone:2011zz} are $\mu_A=m_1$, $\mu_B=m_2$ and $\mu_A=m_A$, $\mu_B=m_B$.

Therefore, one can define the \emph{flavor vacuum} as the state, which is annihilated by flavor annihilation operators at a fixed time $t$ \footnote{Here and throughout we work in the Heisenberg representation.}:
\be
a_{\G k,\si}(t) \, |0(t)\ran_{A,B} \ = \ b_{\G k,\si}(t) \, |0(t)\ran_{A,B} \ = \ 0 \, .
\ee
This is characterized by a boson-condensate structure in terms of modes with definite mass:
\bea 	\label{dcon1}
{}_{A,B}\lan 0(t)|a^\dag_{\G k,1} \, a_{\G k,1}|0(t)\ran_{A,B}  =   {}_{A,B}\lan 0(t)|b^\dag_{\G k,1} \, b_{\G k,1}|0(t)\ran_{A,B} &=&   2 (2 \pi)^3\lf(\cos^2\theta \, |\la^\G k_{1 A}|^2 \, \om_{\G k, A}+\sin^2\theta \, |\la^\G k_{1 B}|^2 \, \om_{\G k, B} \ri) , \\[2mm]
{}_{A,B}\lan 0(t)|a^\dag_{\G k,2} \, a_{\G k,2}|0(t)\ran_{A,B}  =   {}_{A,B}\lan 0(t)|b^\dag_{\G k,2} \, b_{\G k,2}|0(t)\ran_{A,B} &=&   2 (2 \pi)^3\lf(\sin^2\theta \, |\la^\G k_{2 A}|^2 \, \om_{\G k, A} +\cos^2\theta \, |\la^\G k_{2 B}|^2 \, \om_{\G k, B}\ri) , \\[2mm]
{}_{A,B}\lan 0(t)|a^\dag_{\G k,1} \, a_{\G k,2}|0(t)\ran_{A,B}  =   {}_{A,B}\lan 0(t)|b^\dag_{\G k,1} \, b_{\G k,2}|0(t)\ran_{A,B} & = &  
 2 (2 \pi)^3 \sin 2 \theta \, \lf(\la^{\G k*}_{1A} \la^{\G k}_{2A} \, \om_{\G k,A}-\la^{\G k*}_{1B} \la^{\G k}_{2B} \, \om_{\G k,B} \ri)
  . \label{mixcon}
\eea
Shortly we will see that this structure is responsible for the Poincar\'e and $CPT$ symmetry breaking. In particular, the exotic condensates \eqref{mixcon}, which mix particles and antiparticles with different masses could represent a signature of a fundamental dynamical symmetry breaking mechanism that spontaneously breaks Poincar\'e symmetry and at the same time generates mixing (see Refs. \cite{ccpaper,ccpaper3}) in the fermion case. Note that all these condensates vanish for ultrarelativistic modes ($|\G k \gg m_\si|$ and $|\G k \gg m_\si|$). In this regime, eventual effects of SSB should vanish. The same is true also for $\theta=0$.

Flavor states are defined as excitations over the flavor vacuum, i.e.
\be \label{invflavstate}
|a_{\G k,\si}(t)\ran \ \equiv \  a^\dag_{\G k,\si}(t)|0(t)\ran_{A,B}\, , \;\;\;\;\;\;\;\;\; |b_{\G k,\si}(t)\ran \ \equiv \  b^\dag_{\G k,\si}(t)|0(t)\ran_{A,B} \ \, .
\ee
The later are eigenstates of flavor charges
\bea
Q_{\si}(t) & = & i \, \intx \, : \ph^\dag_\si(x) \, \stackrel{\leftrightarrow}{\pa_0} \, \ph_\si(x):
 \ = \  \int \!\! \frac{\dr^3 \G k}{2 \om_{\G k,\si} \, (2 \pi)^3} \, \lf(a^\dag_{\G k,\si}(t) \, a_{\G k,\si}(t)-b^\dag_{\G k,\si}(t) \, b_{\G k,\si}(t)\ri) \, , \label{fc}
\eea
at fixed time\footnote{Here normal ordering is taken with respect to $|0(t)\ran_{A,B}$.} $t$. In particular
\be
Q_{\si}(t) \, |a_{\G k,\si}(t)\ran \ = \ |a_{\G k,\si}(t)\ran\, , \;\;\;\;\;\;\;\;\;   Q_{\si}(t) \, |b_{\G k,\si}(t)\ran \ = \ - |b_{\G k,\si}(t)\ran\, .
\ee
Although flavor charges are not conserved one can introduce the \emph{total flavor charge}:
\be
Q \ = \ \sum_\si \, Q_\si(t) \, ,
\ee
which is conserved ($[Q,H]=0$) and which also satisfies the relation
\be \label{totflav}
Q \, |a_{\G k,\si}(t)\ran \ = \ |a_{\G k,\si}(t)\ran\, , \;\;\;\;\;\;\;\;\;   Q \, |b_{\G k,\si}(t)\ran \ = \ - |b_{\G k,\si}(t)\ran\, .
\ee
From (\ref{fc}) it is also clear that
\begin{eqnarray}
Q_{\si}(t) |0(t)\ran_{A,B} \ = \ Q \,  |0(t)\ran_{A,B} \ = \ 0\, .
\label{32}
\end{eqnarray}
We next proceed to discuss SSB of Poincar\'e symmetry in this system.

%
\section{Spontaneous Poincar\'e symmetry breaking} \label{PoiSSB}
\subsection{Spacetime translations} \label{sttrans}

Let us start by considering spacetime translations, i.e. the subgroup $T^{3,1}$ of the the Poincar\'e group. The generator of {\em space translations} has the usual form:
\be
P_i  \ = \ \sum_{\si=A,B} \, \intx \, \lf(\pi^\dag_\si(x) \pa_i \ph_\si(x) \, + \,  \pi^\dag_\si(x) \, \pa_i \ph^\dag_\si(x) \ri), \qquad i=1,2,3 \, ,
\ee
so that
\be
T(\G b) \ \equiv \ \exp\lf(i \,\G b \cdot \G P\ri) \ = \ \exp\lf(i \,b^i  P_i\ri) \ = \ \exp\lf(-i \,b^i  P^i\ri)\, ,
\ee
and
\be
T(\G b) \, \ph_\si(t,\G x) \,  T^{-1}(\G b) \ = \ \ph_\si(t,\G x+\G b) \, .
\ee
By using the expansion \eqref{expph}, $P_i$ can be rewritten as:
\be
P_i \ = \ \sum_{\si=A,B} \, \int \frac{\dr^3 \G k}{2\om_{\G k,\si}(2 \pi)^3} \, k_i \, \lf(a^\dag_{\G k,\si}(t) \, a_{\G k,\si}(t)+b^\dag_{\G k,\si}(t) \, b_{\G k,\si}(t)\ri) \, .
\ee
This is time independent and commutes with the flavor charge, i.e. $[P_i,Q_\si(t)]=0$ at all times. One can also easily check that
\be
P_i|0(t)\ran_{A,B} \ = \ 0 \, ,
\ee
and so
\be \label{ssym}
T(\G b)|0(t)\ran_{A,B} \ = \ |0(t)\ran_{A,B} \, .
\ee
In other words, {\em flavor vacuum is invariant under space translations}.

The situation changes if one looks at \emph{time translations}. By using canonical commutation relations one can see that
\bea \label{at}
a_{\G k,\si}(t) & = & \sum_{\rho=A,B} \lf(f^\G k_{\si\rho}(t)\, a_{\G k,\rho}(0)+g^\G k_{\si\rho}(t)\, b^\dag_{-\G k,\rho}(0)\ri) \, ,\\[2mm]
b^\dag_{-\G k,\si}(t) & = & \sum_{\rho=A,B} \lf(-g^\G k_{\si\rho}(t)\, a_{\G k,\rho}(0)+f^\G k_{\si\rho}(t)\, b^\dag_{-\G k,\rho}(0)\ri) \, , \label{bt}
\eea
where
\bea \label{fg}
f^\G k_{\si\rho}(t) \ = \ \frac{1}{(2 \pi)^3 \, 2\om_{\G k,\rho}} \, \lf[a_{\G k,\si}(t) \, , \, a^\dag_{\G k,\rho}(0)\ri] \, , \qquad  g^\G k_{\si\rho}(t) \ = \ \frac{1}{(2 \pi)^3 \, 2\om_{\G k,\rho}} \, \lf[b_{-\G k,\rho}(0) \, , \, a_{\G k,\si}(t) \ri] \, .
\eea
The explicit form of these functions is listed in Appendix~B. It is then clear that flavor vacuum is not time-independent. To see this explicitly, let us write the Hamiltonian in the normal-ordered form\footnote{Normal ordering is defined with respect to flavor vacuum at $t=0$.}:
\bea
H \ = \ \intx \, \lf(:\boldsymbol{\pi}^\dag_f(x) \, \boldsymbol{\pi}_f(x)+\nabla \boldsymbol{\ph}^\dag_f (x) \cdot \nabla \boldsymbol{\ph}_f(x)  +\boldsymbol{\ph}^\dag_f(x) \, M^2 \, \boldsymbol{\ph}_f(x) :\ri) \, .
\eea
Because the Hamiltonian is time independent, we can expand it in terms of flavor creation and annihilation operators at $t=0$:
\bea \label{finham}
H & = & \sum_{\si,\tau} \, \int \frac{\dr^3 \G k}{2\om_{\G k,\si}(2 \pi)^3} \,\lf[ w^{\G k}_{\si\tau}\,\lf( a^\dag_{\G k,\si}(0) \, a_{\G k,\tau}(0)+ \, b^\dag_{\G k,\si}(0) \, b_{\G k,\tau}(0) \ri)\ri.\non \\[2mm]
 & + & \lf. y^{\G k}_{\si\tau}\, \lf(a^\dag_{\G k,\si}(0) \, b^\dag_{-\G k,\tau}(0)+b_{-\G k,\si}(0) \, a_{\G k,\tau}(0)\ri) \ri]\, ,
\eea
where the coefficients are given in Eqs.~\eqref{w1}-\eqref{y2}. It is now easy to verify that the Hamiltonian does not annihilate the flavor vacuum, since
\be \label{hoab}
H|0\ran_{A,B} \ = \ \sum_{\si \tau} \,  \int \frac{\dr^3 \G k}{2\om_{\G k,\si}(2 \pi)^3} \, y^{\G k}_{\si \tau} \, |a_{\G k,\si} \ran \otimes | b_{-\G k,\tau}\ran  \ \neq \ 0\, ,
\ee
where $|0\ran_{A,B} \equiv |0(t=0)\ran_{A,B}$. Note, however, that $_{A,B}\langle 0|H|0\ran_{A,B} = 0$ as it should.  Therefore, the {\em symmetry under time translations is spontaneously broken} since the action and ensuing field equations are invariant under time translations. By using Eq.~\eqref{32} one can explicitly verify that the state (\ref{hoab}) carries the zero total charge, i.e.
\be
Q \, H \, |0\ran_{A,B} \ =  \ 0 \, ,
\ee
as we would expect from the conservation of $Q$. We see, therefore, that flavor vacua at different times form a \emph{flavor vacuum manifold}:
\be \label{te}
|0(t)\ran_{A,B} \ = \ T(t)\, |0\ran_{A,B} \, ,
\ee
where
\be
T(t) \ \equiv \ \exp\lf(i \, H \,  t\ri) \, ,
\ee
is the time-evolution operator. The flavor vacuum manifold was introduced in close analogy with \emph{vacuum manifold} defined in the study of SSB in gauge theories. However, here the different vacua are degenerate with respect to total flavor charge and not to energy. In fact, the states representing the flavor vacuum manifold do not posses any sharp value of energy --- energy fluctuates (has a non-trivial variance) on each flavor vacuum~\cite{BJSun}, see also Eq.~(\ref{48bc}).

From Eqs.~\eqref{hoab} and \eqref{te} we can also find that for generic $t$
\be \label{hoabt}
H|0(t)\ran_{A,B} \ = \ \sum_{\si \tau} \,  \int \frac{\dr^3 \G k}{2\om_{\G k,\si}(2 \pi)^3} \, y^{\G k}_{\si \tau} \, |a_{\G k,\si}(t) \ran \otimes | b_{-\G k,\tau}(t)\ran \ \neq \ 0 \, ,
\ee
which completes our proof of the SSB of the time translation symmetry. We have thus proved that spacetime translation symmetry is spontaneously broken on flavor vacuum. The residual vacuum symmetry is then $T^{3}$, i.e. the group of spatial translations.

In passing, we can also establish an analogue of the \emph{Fabri--Picasso theorem}~\cite{FabPic} for the present situation. Let us consider the square norm of $H|0\ran_{A,B}$:
\be
|\!|H|0\ran_{A,B} |\!|^2 \ = \ {}_{A,B}\lan 0|H^2|0\ran_{A,B} \ = \ \intx \, {}_{A,B}\lan 0|H \, T^{00}(\G x)|0\ran_{A,B}  \ = \ \intx \, {}_{A,B}\lan 0|H \, \mathcal{H}(\G x)|0\ran_{A,B} \, ,
\ee
where $T^{00}(\G x)$ and $\mathcal{H}(\G x)$ are the time–time component of energy momentum tensor and Hamilton density, respectively.  Let us regulate $H$ so that
for a sufficiently large space domain $\Omega$ of volume $V$ we introduce $H_V = \int_\Om \dr^3 \G x \ \! \mathcal{H}(\G x)$.
By using the space-translation invariance of the vacuum (cf. Eq.~\eqref{ssym}), we find that
\be
|\!|H_V|0\ran_{A,B} |\!|^2 \ = \ {}_{A,B}\lan 0|H^2_V|0\ran_{A,B} \ = \ V \, {}_{A,B}\lan 0|H_V \, \mathcal{H}(0)|0\ran_{A,B} \, ,
\label{48bc}
\ee
where $V=\int_\Om \dr^3 \G x$. If we now send $V \rightarrow \infty$, we see that (\ref{48bc}) diverges unless $\lim_{V\rightarrow \infty }H_V |0\ran_{A,B} = H|0\ran_{A,B} = 0$. This would, however, be in contradiction with the symmetry breaking condition (\ref{hoab}).
Therefore, the mathematical implementation of these ideas is  rather delicate~\cite{BJV}. The finite volume Hamiltonian
$H_V$ induces a ``finite time translation'', $T_V(t) = \exp(it H_V)$, which in turn gives rise to a ``shifted ground state'', $[|0(t)\ran_{A,B}]_V \ = \ T_V(t)\, |0\ran_{A,B}$. However, very much like the limit $\lim_{V\rightarrow \infty} H_V$ does
not exist, the operator $\exp(it H)$ is not well defined on the flavor Fock space $\mathcal{H}_f(\tau)$ (for any  $\tau$). As a consquence~\cite{BJV}:
\begin{eqnarray}
\lim_{V\rightarrow \infty} {}_{A,B}\langle 0| [|0(t)\ran_{A,B}]_V \ = \ \lim_{V\rightarrow \infty} {}_{A,B}\langle 0| \exp(it H_V) |0\ran_{A,B} \ =  \ 0\, .
\end{eqnarray}
%
%
In other words, flavor Fock spaces at different times are unitarily inequivalent.

The intuitive picture of spontaneous symmetry breaking, based on the observation that a symmetry transformation (\ref{hoab}) does not leave the flavor vacuum state intact, suggests high degeneracy of equivalent flavor vacuum states $|0(t)\ran_{A,B}$. Indeed, since the Hamiltonian $H$  commutes with the charge operator $Q$, so will a finite symmetry transformation $T(t)$ generated by $H$. It will therefore transform the one flavor vacuum state into another
with the same flavor charge. Since the time-translation symmetry group is continuous, we will find infinitely many degenerate flavor vacuum states. On account of the fact that they are all connected by symmetry transformations, they must be physically equivalent and any one of them can serve as a starting point for the construction of the spectrum of excited flavor states. Let us consider, for example, the flavor oscillation formula~\cite{bosons}:
\be \label{oscfor}
\mathcal{Q}_{\si \rightarrow \rho}(t,t_0)\ = \ {}_{A,B}\lan a_{\G k,\si}(t_0)|Q_\rho(t)|a_{\G k,\si}(t_0)\ran_{A,B} \, .
\ee
One can easily verify that
\be
\mathcal{Q}_{\si \rightarrow \rho}(t,t_0) \ = \ \mathcal{Q}_{\si \rightarrow \rho}(t-t_0) \, ,
\ee
i.e. flavor oscillations are \emph{invariant} under time translations. In fact,
\bea
 {}_{A,B}\lan a_{\G k,\si}(t_0)|Q_\rho(t)|a_{\G k,\si}(t_0)\ran_{A,B} & = & {}_{A,B}\lan a_{\G k,\si}(0)|T(t-t_0) \, Q_\rho(0) \, T^{-1}(t-t_0)|a_{\G k,\si}(0)\ran_{A,B}  \non \\[2mm]
& = & {}_{A,B}\lan a_{\G k,\si}(0)|Q_\rho(t-t_0)|a_{\G k,\si}(0)\ran_{A,B} \, ,\label{tinv}
\eea
where we used the group property $T^{-1}(t_0)T(t)=T(t-t_0)$. It is thus clear that the choice of time $t_0$, which we use for the construction of the (Heisenberg representation) state space, is quite immaterial.

 It can also be shown that unlike the transformations of physical states, finite symmetry transformations $T_V(t)$ of observables can be defined consistently in the $V \rightarrow \infty$ limit in theories that are sufficiently causal~\cite{FabPic}. In the following reasonings it will always be implicitly understood that the large-$V$ regulator should be properly employed according to indicated lines whenever expectation values are to be computed.


\subsection{Proper Lorentz group} \label{Sec3}

It is well known that the generator of proper Lorentz algebra $so(3,1)$ can be expressed as \cite{PesSch}
\be
 J^{\mu \nu} \ \equiv \ \intx \, :\lf(x^\mu \, T^{0 \nu}-x^\nu \, T^{0 \mu}\ri): \, , \; \;\;\;\; \,\mu, \nu=0, \dots 3\, .
\ee
Here $T_{\mu \nu}$ is the energy-momentum tensor.

 Let us start from its spatial part:
\be
J_{ij} \ = \ -\sum_{\si} \, \intx \, : x_i \, (\pi^\dag_\si(x) \, \pa_j \ph_\si(x)+\pa_j \ph^\dag_\si(x) \, \pi_\si(x))\, - x_j\,(\pi^\dag_\si(x) \, \pa_i \ph_\si(x)+\pa_i \ph^\dag_\si(x) \, \pi_\si(x)): \, .
\ee
One can equivalently use the angular-momentum operators $J^k$ defined in Eq.~\eqref{identif1}
\be
\G L \ \equiv \ \G J \ = \  - \sum_\si\, \intx \, \lf[\pi^\dag_\si(x) \, (\G x \times \nabla) \, \ph_\si(x) + \ph^\dag_\si(x) \, (\G x \ \!\times \stackrel{\leftarrow}{\nabla}) \, \pi_\si(x) \ri] \, ,
\ee
where we identified $\G J$ with the orbital angular-momentum vector $\G L = (L^1,L^2,L^3)$ because no extra spin contribution is present for scalar fields. In terms of annihilation and creation operators we have:
%
\be
\G L \ = \ \int \frac{\dr^3 \G k}{2\om_{\G k,\si}(2 \pi)^3} \, \lf[a^\dag_{\G k,\si}(t) \, \lf(\G k \times \nabla_\G k \ri) \, a_{\G k,\si}(t) \, + \, b^\dag_{\G k,\si}(t) \, \lf(\G k \times \nabla_\G k \ri) \, b_{\G k,\si}(t)\ri] \, ,
\ee
One can easily verify that
\be
\lf[\G L\, , \, Q_\si(t) \ri] \ = \ \lf[\G L\, , \, H \ri] \ = \ 0 \, .
\ee
It is also not difficult to see that this operator annihilates the flavor vacuum:
\be
\G L \, |0(t)\ran_{A,B} \ = \ 0 \, .
\ee
In fact, we can always perform a unitary canonical transformation which diagonalizes one of the components of the angular momentum\footnote{These cannot be diagonalized simultaneously, because of the $SO(3)$ commutation relations.}. For example, mimicking the case of a free scalar field~\cite{greiner} we can perform the canonical transformation
\bea \label{aplm}
a_{plm,\si}(t) \ \equiv \ i^l \, p \, \int \!\! \dr \Om_p \, Y^*_{lm}(\Om_p) \, a_{\G p,\si}(t) \, , \\[2mm]
b_{plm,\si}(t) \ \equiv \ i^l \, p \, \int \!\! \dr \Om_p \, Y^*_{lm}(\Om_p) \, b_{\G p,\si}(t) \, , \label{bplm}
\eea
where $p=|\G p|$, $Y_{lm}$ are the spherical harmonics and $\Om_p$ is the solid angle at fixed $p$. In this representation $L^3$ has a diagonal form:
\be
L^3 \ = \ \sum_{l,m,\si} \int^\infty_0 \!\! \! \dr p \ m \, \lf(a^\dag_{plm,\si}(t) \, a_{plm,\si}(t)+b^\dag_{plm,\si}(t) \, b_{plm,\si}(t)\ri) \, .
\ee
From Eqs.\eqref{aplm}, \eqref{bplm} it is evident that:
\be
a_{plm,\si}(t)|0(t)\ran_{A,B} \ = \ b_{plm,\si}(t)|0(t)\ran_{A,B} \ = \ 0 \, .
\ee
It follows that $L_3|0\ran \ = \ 0$. The same procedure can be repeated for the other components. In the same way, by defining the generator of rotations
\be \label{rotations}
R(\boldsymbol{\vartheta}) \ = \ \exp\lf(-i \, \boldsymbol{\vartheta} \cdot \G L\ri) \, ,
\ee
one can verify that
\be
R(\boldsymbol{\vartheta})|0(t)\ran_{A,B} \ = \ |0(t)\ran_{A,B} \, ,
\ee
which shows that {\em flavor vacuum is rotationally invariant}.

Let us now analyze the transformation properties  of the flavor vacuum under the Lorentz boosts\footnote{Note that here, as in Ref. \cite{Giacosa:2018dzm} for unstable particles, flavor states have a definite momentum. This is important to remark, because for states that are not energy eigenstates boost and momentum translation are not equivalent.}. The generator of a boost along the $l$-th axis is:
\be
K_l  \ = \ \intx \, :\lf(x^0 \, T^{0l}-x^l \, T^{00}\ri): \, .
\ee
This can also be written as
\be
K_l  \ = \ x^0 \, P^l \, - \, \intx \, x^l \, \mathcal{H}\, .
\ee
In our case:
\bea
K_l & = & \lf[x^0 \, \intx \, :\lf(\boldsymbol{\pi}^\dag_f(x) \, \pa^l \boldsymbol{\ph}_f(x)+\pa^l \boldsymbol{\ph}^\dag_f(x) \, \boldsymbol{\pi}_f(x)\ri):\ri. \non \\[2mm]
& - &  \intx \,x^l :\lf(\boldsymbol{\pi}^\dag_f(x) \, \boldsymbol{\pi}_f(x)+\nabla \boldsymbol{\ph}^\dag_f (x) \cdot \nabla \boldsymbol{\ph}_f(x)  +  \lf. \boldsymbol{\ph}^\dag_f(x) \, M^2 \, \boldsymbol{\ph}_f(x) \ri):\ri] \, .
\label{67bb}
\eea
We can now rewrite (\ref{67bb}) in terms of flavor creation and annihilation operators \eqref{4x4Bog}. By noticing that in the mass basis this is just the sum of boost generators for the two massive fields $\ph_1$ and $\ph_2$ (cf., e.g., Ref.~\cite{ItzZub}), we get
\bea
K_l & = & -i \, \sum_{j=1,2} \, \int \!\!\frac{\dr^3 \G k}{2 \, \om_{\G k,j} (2 \pi)^3} \, :\boldsymbol{a}^\dag_{\G k,j} \, \Om^\G k_j \, \frac{\pa}{\pa k^l} \, \boldsymbol{a}_{\G k,j}:  \\[2mm]
& = & -i \, \sum_{\si,\rho =A,B}\int \!\!\frac{\dr^3 \G k}{2 \, \om_{\G k,\si} (2 \pi)^3} \, :\boldsymbol{a}^\dag_{\G k,\si}(t) \, \Om^{\G k}_{\si\rho}(t) \, \frac{\pa}{\pa k^l} \, \boldsymbol{a}_{\G k,\rho}(t)+\boldsymbol{a}^\dag_{\G k,\si}(t) \, \Om^{\G k}_{\si\rho,l}(t) \, \boldsymbol{a}_{\G k,\rho}(t):\, ,
\eea
where we have have introduced the matrices
\be
\Om^{\G k}_j \ = \ \om_{\G k,j} \, \ide_2 \, , \qquad \Om^{\G k}_{\si\rho} (t) \ = \ \om_{\G k,\si} \, \sum_{j=1,2} \, J^{\G k \dag}_{j\si}(t)  \, J^{\G k}_{j\rho}(t) \, , \qquad  \Om^{\G k}_{\si\rho,l} (t) \ = \  \om_{\G k,\si} \sum_{j=1,2}\, J^{\G k \dag}_{j\si}(t) \, \Om^{\G k}_j \, \frac{\pa}{\pa k^l} \, J^{\G k}_{j\rho}(t) \, ,
\ee
and $\ide_2$ is the $2 \times 2$ identity matrix. The explicit form of $\Om^{\G k}_\si(t)$ and $\Om^{\G k}_{\si,i}(t)$ is not very illuminating and we do not report it here. We only notice that these are non-diagonal matrices.

A generic boost can be thus expressed in the form:
\be
U(L) \ = \ \exp\lf(-i \, \sum^3_{l=1} \, \xi^l \, K_l \ri) \, ,
\ee
where $L(\xi)$ indicates the Lorentz boost transformation:
\be
x^\mu \ \rightarrow \ x'^{\mu} \ = \ L^\mu_{\ \nu}(\xi) \, x^\nu \, .
\ee
%
Now, for flavor fields we can write
\be
U(L) \, \ph_\si(x) \, U^{-1}(L) \ = \ \ph_\si\lf( x'\ri) \, ,
\ee
i.e. $\ph_\si$ behaves as a scalar under Lorentz boost. From Eq.~\eqref{expph} we get:
\bea
U(L) \, \ph_\si(x) \, U^{-1}(L) & = & \int \frac{\dr^3 \G k}{2\om_{\G k,\si}(2 \pi)^3} \, \lf[a_{\G k,\si}\lf(t'\ri)\, e^{-i k \, x'}\ + \ b^{\dagger}_{\G k,\si}\lf(t'\ri) \, e^{i k \, x'}\ri] \, .
\eea
Here and in the following we formally use the notation $L \G k$ to indicate $L^j_\mu k^\mu$ (j=1,2,3), respectively. This equation should be actually written in the form:
\bea
U(L) \, \ph_\si(x) \, U^{-1}(L) & = & \int \!\! \frac{\dr^4 k}{(2\pi)^4} \, (2\pi) \, \de^4(k^2-\mu_\si^2) \theta(k_0) \, \lf[a_{k,\si}\lf(t'\ri)\, e^{-i k \, x'}\ + \ b^{\dagger}_{k,\si}\lf(t'\ri) \, e^{i k \, x'}\ri] \, .
\eea
Performing the change of variables~\cite{PesSch}: $k \rightarrow  k'=L^{-1} k$, we have:
\bea
U(L) \, \ph_\si(x) \, U^{-1}(L) & = & \int \!\! \frac{\dr^4 k}{(2\pi)^4} \,(2\pi) \, \de^4(k^2-\mu_\si^2) \theta(k_0) \, \lf[a_{L k,\si}\lf(t'\ri)\, e^{-i k x}\ + \ b^{\dagger}_{L k,\si}\lf(t'\ri) \, e^{i k \, x}\ri] \, .
\eea
By integrating over $k_0$ we find:
\bea
U(L) \, \ph_\si(x) \, U^{-1}(L) & = & \int \frac{\dr^3 \G k}{2\om_{\G k,\si}(2 \pi)^3} \, \lf[a_{L \G k,\si}\lf(t'\ri)\, e^{-i k \, x} \ + \ b^{\dagger}_{L \G k,\si}\lf(t'\ri) \, e^{i k \, x}\ri] \, .
\eea
Therefore, by comparing with Eq.\eqref{expph} we find
\bea   \label{lt1}
U(L) \, a_{\G k,\si}(t) \, U^{-1}(L) & = & a_{L \G k,\si}\lf(t'\ri) \, , \\[2mm]
U(L) \, b_{\G k,\si}(t) \, U^{-1}(L) & = & b_{L \G k,\si}\lf(t'\ri) \, . \label{lt2}
\eea
To find the explicit form of these operators, in terms of the ones at time $t$, we can employ canonical commutation relations to get:
\bea
&&\hspace{-0.5cm} U(L) \, a_{\G k,\si}(t) \, U^{-1}(L) \ = \  a_{L \G k,\si}\lf(t'\ri) \\[2mm] \non
&&\hspace{-0.5cm} = \ \sum_{\rho=A,B}\frac{1}{2 \om_{L \G k;\rho}(2\pi)^3}\lf(\lf[a_{L \G k,\si}\lf(t'\ri) \, , \, a^\dag_{L \G k,\rho}(t)\ri] \, a_{L \G k,\rho}(t) \, - \,\lf[a_{L \G k,\si}\lf(t'\ri) \, , \, b_{-L \G k,\rho}(t)\ri] \, b^\dag_{-L \G k,\rho}(t)\ri) \, ,
\eea
and similar relations hold also for the other operators\footnote{Note that commutators at different times are $c$-numbers.}. These are analogous to Eqs.~\eqref{at}-\eqref{bt}. If  we now look at flavor-vacuum transformation properties under boosts we have
\bea
\mbox{\hspace{-8mm}}&& |0\lf(t';\boldsymbol{\xi} \ri)\ran_{A,B} \ = \ U(L)|0(t)\ran_{A,B}\non \\[2mm]
\mbox{\hspace{-8mm}}&& = \  \exp\lf(-\sum^3_{l=1}\xi^l  \sum_{\si,\rho =A,B}\int \!\!\frac{\dr^3 \G k}{2 \, \om_{\G k,\si} (2 \pi)^3} \, :\boldsymbol{a}^\dag_{\G k,\si}(t) \, \Om^{\G k}_{\si\rho}(t) \, \frac{\pa}{\pa k^l} \, \boldsymbol{a}_{\G k,\rho}(t)+\boldsymbol{a}^\dag_{\G k,\si}(t) \, \Om^{\G k}_{\si\rho,l}(t) \, \boldsymbol{a}_{\G k,\rho}(t):\ri) \, |0(t)\ran_{A,B} \, .
\label{boostvac}
\eea
We see immediately that $|0\lf(t';\boldsymbol{\xi} \ri)\ran_{A,B} \neq  |0(t)\ran_{A,B}$ and hence the flavor vacuum  is changed under the action of Lorentz boosts, while the action and ensuing field  equations  stay unchanged. In other words, the {\em Lorentz boosts symmetry is spontaneously broken on flavor vacuum}. By varying $\boldsymbol{\xi}$, we generate a flavor vacuum manifold of unitarily inequivalent states, as in  the case of flavor vacua at different times, which we analyzed in Section~\ref{sttrans}. In fact, since
\be\label{boosts}
[K_l \, , \, Q] \ = \ 0 \, ,
\ee
all states defined in~\eqref{boostvac}  correspond to zero total flavor  charge. Same considerations as in the time-translation case, based on Fabri--Picasso theorem and
the large-$V$ regularization, can be repeated here in the same way.

From the above discussion it is clear that only rotational symmetry $SO(3)$, whose generators are given by Eq.~\eqref{rotations} is a symmetry of the flavor vacuum. This result, together with the one of the previous section, tells us that the flavor vacuum symmetry group is the Euclidean group $E(3)$, as stated in Section~\ref{pomix}. The quadratic Casimir of this group are~\cite{Tung} $\G P^2 \ \equiv \ \G P \cdot \G P$ and $\G J \cdot \G P$, which now substitute $P^2$ and $W^2$. It is worthy of remarking that similar results were derived in the case of unstable particles~\cite{DeFVit,Exener}, which strengthens even more the analogy between flavor mixing and unstable particles proposed in Ref.~\cite{BJSun}.
Note that the flavor vacuum manifold has together 4  flavor flat directions (i.e. directions along which the total flavor charge remains zero) corresponding to the number of broken generators. In particular, the flavor vacuum manifold ${\mathcal{M}} = \left\{ |0\lf(t;\boldsymbol{\xi} \ri)\ran_{A,B}, (t,\boldsymbol{\xi}) \in \mathbb{R}^4  \right\}$ is isomorphic to the quotient space $(T^{3,1} \rtimes \, O(3,1))/E(3)$. Note that dimension of the quotient space, i.e.,  $\dim[(T^{3,1} \rtimes \, O(3,1))/E(3)]$ is correctly $10-6 =4$. 
Let us also observe that there are no {\em energy flat directions} on ${\mathcal{M}}$.  Indeed, from the Fabri--Picasso  theorem [cf. Eq.~(\ref{48bc})] we see
that  the variance of the energy is infinite at any point on the vacuum manifold $\mathcal{M}$, which in turn prohibits the existence of energy flat directions on $\mathcal{M}$. Note that such a divergence  is basically an infrared problem (large-$V$ problem) and  it can be controlled by means of an appropriate regularization scheme.
This argument indicates that there should be {\em no Goldstone modes} present in the theory, since these are normally associated with gapless fluctuations along flat energy directions.

So, while the charge $Q_A(t)$  {\em does not} fluctuate on the state  $|0\lf(t;\boldsymbol{\xi} \ri)\ran_{A,B} \in \mathcal{M}$, the fluctuations of $E$ are on the very same state {\em unbounded}.  This complementarity between $E$ and $Q_{A}$ fluctuations on $\mathcal{M}$ might also be viewed as a direct manifestation of flavor-energy uncertainty relations~\cite{BJSun}\footnote{
In fact,
for any label time $\tau$ there exists $Q_{\si}(\tau)$ such that $Q_{\si}(\tau)|0\lf(\tau;\boldsymbol{\xi} \ri)\ran_{A,B} = 0$ (cf. Eq.~(\ref{32})) but $[Q_{\si}(t), H] \neq 0$.
Let us now consider the flavor-energy uncertainty relations~\cite{BJSun}
\be
\De E \, \De Q_\si(t) \ \geq \ \frac{1}{2}\lf|\frac{\dr \lan Q_{\si}(t)\ran}{\dr t}\ri| \, ,
\label{neutun}
\ee
where $\De Q_\si$ and $\De E$ are standard deviations of charge and energy, respectively
evaluated on $|0\lf(\tau;\boldsymbol{\xi} \ri)\ran_{A,B}$ flavor vacuum at the fixed label time (e.g. $\tau=0$). Because $Q_{\si}(0)|0\lf(0;\boldsymbol{\xi} \ri)\ran_{A,B} = 0$
we have that $\De Q_\si(t)|_{\tau\rightarrow 0} = 0$. 
The RHS of (\ref{neutun}) equals zero only for $\theta=0$ or $m_1=m_2$, i.e., for the non-mixing case. This is, however, trivial situation since in this  case  $|0\ran_{A,B}=|0\ran_{1,2}$ and hence no symmetry breaking is present.
On the other hand, for $\theta \neq 0$, the RHS of (\ref{neutun}) is non-zero, while on the LHS $\De Q_\si(t)|_{t\rightarrow 0} = 0$, implying $\Delta E \to \infty$. 
}.

As in the case of time-translation, we can now show that different states in the flavor vacuum manifold are physically equivalent. In other words, flavor oscillations can be equivalently described in every Lorentz frame. Let us consider a flavor wavepacket:
\be
|a_{\si}(y)\ran \ \equiv \ \int \frac{\dr^3 \G k}{2\om_{\G k,\si}(2 \pi)^3} \, e^{-i k y} \,f(\G k)\, a_{\G k,\si}(y_0)|0(y_0)\ran_{A,B} \, ,
\ee
and suppose that the momentum space distribution $f(\G k)$ is Lorentz invariant. Therefore:
\be
|a'_{\si}(y)\ran \ \equiv \ U(L) |a_{\si}(y)\ran \ = \ |a_{\si}\lf(y' \ri)\ran \, ,
\ee
as one can derive from Eqs.~\eqref{lt1},\eqref{lt2}. Covariant oscillation formula should be written as:
\be \label{covosc}
\mathcal{J}^\mu_{\si\rightarrow \rho}(x-y)\ = \ \lan a_{\si}(y)|J^\mu_\rho(x)|a_{\si}(y)\ran \, ,
\ee
where $J^\mu_\rho(x)$ are the flavor currents defined as~\cite{bosons}
\be\label{fcur}
J^\mu_\rho(x) \ \equiv \ i \ph^\dag_\rho(x) \stackrel{\leftrightarrow}{\pa^\mu} \ph_\rho(x) \, .
\ee
Clearly, Eq.~\eqref{oscfor} can be obtained by taking $\mu=0$ and integrating on space variables.

In the primed Lorentz frame Eq.\eqref{oscfor} reads
\be \label{covoscp}
\lan a'_{\si}(y)|J^\mu_\rho(x')|a'_{\si}(y)\ran \ = \ \lan a_{\si}(y')|J^\mu_\rho(x')|a_{\si}(y')\ran \ = \ \mathcal{J}^\mu_{\si\rightarrow \rho}(x'-y') \, .
\ee
Therefore, the flavor oscillation formula in the primed Lorentz frame is the same as in the unprimed one. This shows, once more, that Poincar\'e (and Lorentz) symmetry  breaking on the flavor vacuum, which leads to
non-zero vector current vacuum expectation values~\eqref{fcur}, has no direct consequences on flavor oscillations.  Thus, Poincar\'e  invariance breaking contributions to QFT oscillation formula as reported in~\cite{BLDiMLa} are  mere artifacts of the non-covariant formalism (oscillations in time) used in that work.

As we have seen above, another important feature of the  Poincar\`e/Lorentz SSB via the dynamical flavor condensates is the apparent {\it absence}
of any Goldstone bosons, as discussed above.  Thus the spectrum of the flavor vacuum  remains the same as the mass eigenstate one, and we have no extra massless modes. This situation is to be contrasted
with the standard lore of non-flavored QFT. Indeed, it has been suggested in~\cite{tomboulis}, that, in gauge theories with Lorentz SSB, in the sense of a vector gauge boson acquiring a vacuum expectation value, the massless U(1) photon plays the r\^ole of such a Goldstone boson. In the current, non gauge, context, although the flavor currents~\eqref{fcur}
acquire non-zero vacuum expectation values~\eqref{covoscp} in terms of the flavor vacuum, nonetheless, as we explained above, they are {\it not associated with any} Goldstone bosons.

\section{Discrete symmetries} \label{Sec4}

Until now we did not consider the discrete symmetries. However, they have to be included in a complete study of Lorentz group properties of flavor operators. Moreover, in the current literature, Lorentz symmetry breaking is often discussed in parallel with $CPT$ symmetry breaking~\cite{Colladay:1996iz,cptgreen,Kostelecky2004}, because the $CPT$ theorem strongly depends on the assumption of Lorentz invariance~\cite{StWi}.

In this section we study the behavior of flavor annihilation and creation operators  under parity, charge conjugation and time reversal. This will be done by considering discrete symmetries both separately and in different relevant combinations. We will see that time reversal is spontaneously broken and as consequence also $CPT$ is not a symmetry of the flavor vacuum.

\subsection{Parity}

The parity transformation of the flavor scalar fields is given by:
\be
P \, \ph_\si(x) \, P^{-1} \ = \ \eta_{\si,{}_{P}} \, \ph_\si(\tilde{x}) \, ,
\ee
where $P$ is the unitary parity operator and $\tilde{x}=(t,-\G x)$. We choose the intrinsic parity to satisfies $|\eta_{\si,{}_{P}}|^2=1$. By using the explicit expansion \eqref{expph}, we find:
\bea
P \, \ph_\si(x) \, P^{-1} & = & \int \frac{\dr^3 \G k}{2\om_{\G k,\si}(2 \pi)^3} \, \lf[P \, a_{\G k,\si}(t) \, P^{-1}\, e^{-i\om_{\G k,\si} t}\ + \ P \, b^{\dagger}_{-\G k,\si}(t) \, P^{-1} \, e^{i\om_{\G k,\si} t}\ri] \, e^{i\G k \cdot \G x} \non \\[2mm]
& = &  \eta_{\si,{}_{P}} \, \int \frac{\dr^3 \G k}{2\om_{\G k,\si}(2 \pi)^3} \, \lf[a_{\G k,\si}(t) \, e^{-i\om_{\G k,\si} t}\ + \ b^{\dagger}_{-\G k,\si}(t) \, \, e^{i\om_{\G k,\si} t}\ri] \, e^{-i\G k \cdot \G x} \, .
\eea
Consequently, transformations of creation and annihilation operators satisfy the following relations:
\bea
P \, a_{\G k,\si}(t) \, P^{-1} & = & \eta_{\si,{}_{P}} \, a_{-\G k,\si}(t) \, , \qquad P \, b_{\G k,\si}(t) \, P^{-1} \ = \ \eta^*_{\si,{}_{P}} \, b_{-\G k,\si}(t) \, , \\[2mm]
P \, a^\dag_{\G k,\si}(t) \, P^{-1} & = & \eta^*_{\si,{}_{P}} \, a^\dag_{-\G k,\si}(t) \, , \qquad P \, b^\dag_{\G k,\si}(t) \, P^{-1} \ = \ \eta_{\si,{}_{P}} \, b^\dag_{-\G k,\si}(t) \, .
\eea
It can be checked that the explicit form of $P$  satisfying above relations reads (see also Ref.~\cite{greiner})
\bea
P & = & \exp\lf\{i \frac{\pi}{2} \, \int \frac{\dr^3 \G k}{2\om_{\G k,\si}(2 \pi)^3} \, \lf[ \lf(a^\dag_{-\G k,\si}(t) \, a_{\G k,\si}(t)+b^\dag_{-\G k,\si}(t) \, b_{\G k,\si}(t)\ri) \ri. \ri. \non \\[2mm]
& - &\lf. \lf.  \eta_{\si,{}_{P}}  \lf(a^\dag_{\G k,\si}(t) \, a_{\G k,\si}(t)+b^\dag_{\G k,\si}(t) \, b_{\G k,\si}(t)\ri) \ri]\ri\} \, .
\eea
%
By inspection we see that the {\em flavor vacuum is invariant under parity transformation}, i.e., up to an irrelevant phase factor we have
\be \label{pifv}
P \, |0(t)\ran_{A,B} \ = \ |0(t)\ran_{A,B} \, .
\ee
As a simple consequence we get that
\be
P\, |a_{\G k,\si}(t)\ran \ = \ |a_{-\G k,\si}(t)\ran \, ,
\ee
and flavor charges \eqref{fc} remain invariant, i.e.
\be
\lf[P \, , \, Q_\si(t)\ri] \ = \ 0 \, .
\ee
%
\subsection{Charge conjugation}

The charge conjugation transformation of the flavor scalar fields is given by
\be
C \, \ph_\si(x) \, C^{-1} \ = \ \eta_{\si,{}_{C}} \, \ph^\dag_\si(x) \, ,
\ee
where $C$ is the unitary charge conjugation operator. Again, our convention is $|\eta_{\si,{}_{C}}|^2=1$. Once more, by using the explicit expansion \eqref{expph}, we find:
\bea
C \, \ph_\si(x) \, C^{-1} & = & \int \frac{\dr^3 \G k}{2\om_{\G k,\si}(2 \pi)^3} \, \lf[C \, a_{\G k,\si}(t) \, C^{-1}\, e^{-i \, k \, x}\ + \ C \, b^{\dagger}_{\G k,\si}(t) \, C^{-1} \, e^{i \, k \, x}\ri] \non \\[2mm]
& = &  \eta_{\si,{}_{C}} \, \int \frac{\dr^3 \G k}{2\om_{\G k,\si}(2 \pi)^3} \, \lf[a^\dag_{\G k,\si}(t) \, e^{i \, k \, x}\ + \ b_{\G k,\si}(t) \, \, e^{-i \, k \, x}\ri] \, .
\eea
Transformations of creation and annihilation operators follow:
\bea
C \, a_{\G k,\si}(t) \, C^{-1} & = & \eta_{\si,{}_{C}} \, b_{\G k,\si}(t) \, , \qquad C \, b_{\G k,\si}(t) \, C^{-1} \ = \ \eta^*_{\si,{}_{C}} \, a_{\G k,\si}(t) \, , \\[2mm]
C \, a^\dag_{\G k,\si}(t) \, C^{-1} & = & \eta^*_{\si,{}_{C}} \,  b^\dag_{\G k,\si}(t) \, , \qquad C \, b^\dag_{\G k,\si}(t) \, C^{-1} \ = \ \eta_{\si,{}_{C}} \, a^\dag_{\G k,\si}(t) \, .
\eea
From this, the explicit form of $C$ reads
\bea
C & = & \exp\lf\{i \frac{\pi}{2} \, \int \frac{\dr^3 \G k}{2\om_{\G k,\si}(2 \pi)^3} \, \lf[ \lf(b^\dag_{\G k,\si}(t) \, a_{\G k,\si}(t)+a^\dag_{\G k,\si}(t) \, b_{\G k,\si}(t)\ri) \ri. \ri. \non \\[2mm]
& - &\lf. \lf.  \eta_{\si,{}_{C}}  \lf(a^\dag_{\G k,\si}(t) \, a_{\G k,\si}(t)+b^\dag_{\G k,\si}(t) \, b_{\G k,\si}(t)\ri) \ri]\ri\} \, ,
\eea
which shows that the {\em flavor vacuum is invariant under charge conjugation}, i.e.
\be \label{ccfv}
C \, |0(t)\ran_{A,B} \ = \ |0(t)\ran_{A,B} \, .
\ee
Consequently,  a flavor state~\eqref{invflavstate} transforms as
\be
C\, |a_{\G k,\si}(t)\ran \ = \ |b_{\G k,\si}(t)\ran \, ,
\ee
while flavor charge \eqref{fc} reverses its sign
\bea
C \, Q_{\si}(t) \, C^{-1} \ = \ \int \!\!
\frac{\dr^3 \G k}{2 \om_{\G k,\si} 	 (2 \pi)^3}
\, \lf(b^\dag_{\G k,\si}(t) \,
b_{\G k,\si}(t)-a^\dag_{\G k,\si}(t) \, a_{\G k,\si}(t)\ri)\ = \ - Q_{\si}(t)\, ,
\eea
as expected.
\subsection{Time reversal}
The time reversal transformation of the flavor scalar fields is given by:
\be
T \, \ph_\si(x) \, T^{-1} \ = \ \eta_{\si,{}_{T}} \, \ph_\si(-\tilde{x}) \, ,
\ee
where $T$ is the \emph{antiunitary} time reversal operator. We employ the convention for the phase $|\eta_{\si,{}_{T}}|^2=1$. By using the explicit expansion \eqref{expph}, we find:
\bea
T \, \ph_\si(x) \, T^{-1} & = & \int \frac{\dr^3 \G k}{2\om_{\G k,\si}(2 \pi)^3} \, \lf[T \, a_{\G k,\si}(t) \, T^{-1}\, e^{i\om_{\G k,\si} t}\ + \ T \, b^{\dagger}_{-\G k,\si}(t) \, T^{-1} \, e^{-i\om_{\G k,\si} t}\ri] \, e^{-i\G k \cdot \G x} \non \\[2mm]
& = &  \eta_{\si,{}_{T}} \, \int \frac{\dr^3 \G k}{2\om_{\G k,\si}(2 \pi)^3} \, \lf[a_{\G k,\si}(-t) \, e^{i\om_{\G k,\si} t}\ + \ b^{\dagger}_{-\G k,\si}(-t) \, \, e^{-i\om_{\G k,\si} t}\ri] \, e^{i\G k \cdot \G x} \, .
\eea
Transformations of creation and annihilation operators follow:
\bea
T \, a_{\G k,\si}(t) \, T^{-1} & = & \eta_{\si,{}_{T}} \, a_{-\G k,\si}(-t) \, \qquad T \, b_{\G k,\si}(t) \, T^{-1} \ = \ \eta^*_{\si,{}_{T}} \, b_{-\G k,\si}(-t) \, , \\[2mm]
T \, a^\dag_{\G k,\si}(t) \, T^{-1} & = & \eta^*_{\si,{}_{T}} \, a^\dag_{-\G k,\si}(-t) \, \qquad T \, b^\dag_{\G k,\si}(t) \, T^{-1} \ = \ \eta_{\si,{}_{T}} \, b^\dag_{-\G k,\si}(-t) \, .
\eea
Let us note in this connection that for flavor $A$ we can explicitly write

\bea
&& T \, a_{\G k,A}(t) \, T^{-1} \ = \eta_{\si,{}_{T}} \, a_{-\G k,A}(-t)  \\[2mm] \non
&& = \ \eta_{\si,{}_{T}} \, \sum_{\rho=A,B} \frac{1}{2 \om_{\G k,\rho}(2\pi)^3}\lf(\lf[a_{-\G k,\si}(-t) \, , \, a^\dag_{ -\G k,\rho}(t)\ri] \, a_{-\G k,\rho}(t) - \lf[a_{-\G k,\si}(-t) \, , \, b_{\G k,\rho}(t)\ri] \, b^\dag_{\G k,\rho}(t)\ri) \, ,
\eea
where on the second line the result is phrase in terms of operators  $a_{-\G k,\rho}(t)$ and $b^\dag_{\G k,\rho}(t)$ at original time $t$. Commutators involved are just c-numbered functions due to a quadratic nature of our model system.   Similar relations hold for the other operators and flavor $B$. If one now looks at flavor vacuum transformation properties
\be \label{tsb}
|0(t)\ran^{T}_{A,B} \ = \ T|0(t)\ran_{A,B} \, ,
\ee
one finds  that time-reversal symmetry is spontaneously broken. This could also be seen by looking at flavor charge \eqref{fc} transformation:
\bea
T \, Q_{\si}(t) \, T^{-1} \ = \ \int \!\! \frac{\dr^3 \G k}{2 \om_{\G k,\si} 	 (2 \pi)^3} \, \lf(a^\dag_{\G k,\si}(-t) \, a_{\G k,\si}(-t)-b^\dag_{\G k,\si}(-t) \, b_{\G k,\si}(-t)\ri)\ = \ Q_{\si}(-t)\, ,
\eea
i.e., $[Q_\si(t),T] \neq 0$ in a non-trivial way (they neither commute or anticommute). This implies that
\begin{eqnarray}
Q_\si(t)T |0(t)\ran_{A,B} \ = \ Q_\si(t) |0(t)\ran^{T}_{A,B} \ \neq \ 0\, ,
\end{eqnarray}
while $Q_\si(t)|0(t)\ran_{A,B} = 0$. This shows that the {\em time-reversal symmetry is spontaneously broken}.

Once more, we notice that oscillation formula for our toy-model system is left unchanged by time reversal transformation. In fact, from Eq.~\eqref{oscfor}, we have
\be \label{trinv}
\mathcal{Q}_{\si \rightarrow \rho}(-t) \ = \ \lan a_{\G k,\si}(0)|{Q}_\rho(-t)|a_{\G k,\si}(0)\ran \ = \ \lan a_{\G k,\si}(0)|T \, {Q}_\rho(t) \, T^{-1} |a_{\G k,\si}(0)\ran \ = \ \mathcal{Q}_{\si \rightarrow \rho}(t) \, ,
\ee
where we used that
\be
T^{-1} |a_{\G k,\si}(0)\ran \ = \  |a_{\G k,\si}(0)\ran \, .
\ee
%
\subsection{CP and CPT symmetry}
From the previous considerations it is evident that $CP$ is an exact symmetry in the flavor representation \footnote{This is not true for the three flavor case, where $CP$ symmetry can be \emph{explicitly} broken because of a complex phase in the mass matrix.}:
\be
C \, P \, |0(t)\ran_{A,B} \ = \  |0(t)\ran_{A,B} \, .
\ee
However, from Eq.\eqref{tsb}, it follows that $CPT$ symmetry is spontaneously broken on the flavor vacuum:
\be \label{cptsb}
|0(t;\Theta)\ran_{A,B} \ = \ \Theta \, |0(t)\ran_{A,B} \, ,
\ee
where $\Theta\equiv C \, P\, T$. This is a consequence of the transformation law of creation and annihilation operators:
\bea
\Theta \, a_{\G k,\si}(t) \, \Theta^{-1} & = & \eta_{\si} \, b_{\G k,\si}(-t) \, , \qquad \Theta \, b_{\G k,\si}(t) \, \Theta^{-1} \ = \ \eta^*_{\si} \, a_{\G k,\si}(-t) \, , \\[2mm]
\Theta \, a^\dag_{\G k,\si}(t) \, \Theta^{-1} & = & \eta^*_{\si} \, b^\dag_{\G k,\si}(-t) \, , \qquad \Theta \, b^\dag_{\G k,\si}(t) \, \Theta^{-1} \ = \ \eta_{\si} \, a^\dag_{\G k,\si}(-t) \, .
\eea
where $\eta_{\si}\equiv \eta_{\si,{}_{C}} \eta_{\si,{}_{P}} \eta_{\si,{}_{T}}$. This implies the charge transformation:
\bea \non
\Theta \, Q_{\si}(t) \, \Theta^{-1} & = & \int \!\! \frac{\dr^3 \G k}{2 \om_{\G k,\si} 	 (2 \pi)^3} \, \lf(b^\dag_{\G k,\si}(-t) \, b_{\G k,\si}(-t)-a^\dag_{\G k,\si}(-t) \, a_{\G k,\si}(-t)\ri)\ = \ -Q_{\si}(-t)\, .
\eea
By repeating the same reasoning as in Section~IV we obtain for the flavor current
\be \label{CPTinv}
\mathcal{J}^\mu_{\si \rightarrow \rho}(x) \ = \ \mathcal{J}^\mu_{\bar{\si} \rightarrow \bar{\rho}}(-x) \, ,
\ee
i.e. {\em flavor oscillations are $CPT$ invariant}.

\section{Conclusions and Outlook}\label{conclusions}

In this paper, we have studied the non-trivial behavior of flavor states with respect to Poincar\'e and $C$,$P$ and $T$ symmetry and we argued that flavor states are not compatible with Poincar\'e symmetry. Instead of extending Poincar\'e, as proposed in Ref.~\cite{Lobanov}, we show that the flavor Fock space constructed {\em \`{a} la} Refs.~\cite{BlaVit95,qftmixing,Binger:1999nj,bosons}, naturally leads to Poincar\'e SSB, with the residual symmetry of the vacuum state being $E(3)$. This SSB is caused by the non-trivial flavor condensate structure [see Eqs.\eqref{dcon1}-\eqref{mixcon}], which, however, becomes phenomenologically insignificant for ultra-relativistic modes and also for mixing angle $\theta=0$.

In order to demonstrate our point, we analyzed the properties of flavor creation and annihilation operators under Poincar\'e and discrete symmetry transformations, in a toy-model describing a flavor scalar doublet with mixing. Moreover, we have defined \emph{flavor vacuum manifold} as the set of flavor-degenerate states (all with zero-flavor charge). We have provided explicit examples of flavor vacua at label times, and  in different Lorentz frames. With the help of the Fabri--Picasso theorem we showed that the respective flavor Fock spaces are unitarily inequivalent. We also proved that time-reversal and $CPT$ symmetries are spontaneously broken, while $CP$ symmetry is exact, in our two-flavor case, as expected. However, this type of SSB of Poincar\'e and $CPT$ symmetry does not imply the presence of any Goldstone bosons or Poincar\'e or CPT violating effects in the flavor oscillations formula, which is of phenomenological interest.

Nonetheless, we should remark at this stage that the
flavor-vacuum energy term, associated with the Lorentz- and $CPT$-violating flavor condensate, might have other non-trivial phenomenological consequences, when the model is properly extended to cosmology. Indeed, it is known~\cite{cosmo}, that the non-perturbative condensate of flavor-vacua leads to novel contributions to dark energy. Our current work points to the fact that such contributions break spontaneously the Lorentz and $CPT$ symmetries of the Universe ground state.  It would then be interesting to study the effects of such flavor-induced Lorentz- and $CPT$-violating  effects ({cf.} the vector
vacuum expectation value~\eqref{fcur}) on the early Universe, such as their imprint on  cosmic microwave background, inflationary perturbations,  etc..

It should be stressed that our analysis is related to the problem of dynamical mixing generation~\cite{mavroman,DynMix,ccpaper,ccpaper3}. In fact, in such a context one can explain the origin of Poincar\'e and $CPT$ symmetry breaking together with the origin of field mixing. In this direction, another interesting possibility is that such a mechanism, {when properly extended to chiral fermions,
could lead, through the Lorentz- and $CPT$- violating flavor-vacuum chiral condensates,
to phenomena like the \emph{chiral magnetic effect}~\cite{chiral} in quantum chromodynamics: the Lorentz violating condensate on flavor vacuum can act as a finite temperature background, where a current $J$ is dynamically generated in regions with an external magnetic field. We reserve a further detailed study of such speculative issues for a future work.

\section*{Acknowledgments} L.S. would like to thank F.~Iachello for useful comments.
The work of P.J.  was  supported  by the Czech  Science  Foundation Grant No. 19-16066S, while that of NEM is supported in part by the UK Science and Technology Facilities  research Council (STFC) under the research grants
ST/P000258/1 and ST/T000759/1, and by the COST Association Action CA18108 ``{\it Quantum Gravity Phenomenology in the Multimessenger Approach (QG-MM)}''.
NEM also acknowledges a scientific associateship (``\emph{Doctor Vinculado}'') at IFIC-CSIC-Valencia University, Valencia, Spain.
\appendix
\section{Basic structure of the Poincar\'e group}

In order to fix the notation and the conventions, we briefly review the main features of Lorentz and Poincar\'e group, following Ref.~\cite{Tung}. Given the Minkowski space $\lf(\mathbb{R}^4,\dr s^2\ri)$ where $ds^2$ is the indefinite quadratic form:
\be \label{Minkquad}
\dr s^2=g_{\mu \nu} \dr x^\mu \dr x^ \nu\, ,
\ee
and $g=\mathrm{diag}(1,-1,-1,-1)$ is the metric tensor.

The \emph{homogeneous Lorentz group} is the set of transformations which leave unchanged the quadratic form \eqref{Minkquad}. This definition can be expressed from the relation:
\be \label{Lordef}
g_{\mu \nu} \Lambda^\mu_\la \Lambda^\nu_\si=g_{\la \si} \, .
\ee
Because of the symmetry of the metric tensor these are 10 independent constraints. Therefore, the Lorentz group has six parameters. If in Eq.~\eqref{Lordef} we put $\la=\si=0$ we find the condition
\be
\lf(\La^0_0 \ri)^2-\sum^3_{i=1} \lf(\La^i_0 \ri)^2=1 \, ,
\ee
and then, $\lf(\La^0_0 \ri)^2 \geq 0$, i.e. $\La^0_0 \geq 0$ or $\La^0_0 \leq 0$. Considering only the transformations continuously connected with the identity we must choose only the first condition. Moreover
\be
\lf(\mathrm{det} \La \ri)^2=1 \, .
\ee
Because we are limiting ourselves to transformations that are continuously connected with the identity, we must consider only the case $\mathrm{det} \La=1$. These two choices define the \emph{proper orthochronous Lorentz group} $SO^{+}_{\uparrow}(3,1)$. If these restrictions are dropped (e.g., when discrete $P$ and $T$ symmetries are included) one speaks about the {\em full Lorentz group}.

The spatial part of Eq.~\eqref{Lordef} can be rewritten as the condition
\be
R^{-1} \ = \ R^T \, ,
\ee
that defines the group of $O(3)$ matrices. The condition on the determinant is fulfilled by $SO(3)$ matrices which thus define a three parameters subgroup of the proper Lorentz group.  A second large (3-parametric) class of Lorentz transformations consists of the so-called \emph{Lorentz boosts} (or special Lorentz transformations). These represent class of rotation-free Lorentz transformation.
The Lorentz boosts do not form a group --- successive boosts along non-parallel directions do not yield a boost, but the combination of a boost and spatial rotation.  For instance, a Lorentz boost along the $x$ axis is of the form:
\be \label{L1}
L_1 \ = \ \bbm
\cosh \xi && \sinh \xi && 0 && 0 \\
\sinh \xi && \cosh \xi && 0 && 0 \\
0 && 0 && 1 && 0 \\
0 && 0 && 0 && 1
\ebm \, .
\ee
This represents the transformation between an inertial frame and another inertial frame, moving along the $x$ axis with velocity $v=c \tanh \xi$. The parameter $\xi$ is known as {\em rapidity} and since $-c \leq v  \leq c$ one has that $-\infty<\xi<+\infty$, so the full Lorentz group, which is indicated with $SO(3,1)$, is non-compact. One can also prove that a general Lorentz transformation within $SO^{+}_{\uparrow}(3,1)$ can be decomposed in terms of boosts and rotations as:
\be
\La=R(\al, \bt, 0) L_3(\xi) R(\phi, \theta,\psi)^{-1} \, ,
\ee
where the rotation matrix are written in terms of Euler's angles.

The \emph{inhomogeneous Lorentz group} or \emph{Poincar\'e group}, includes also spacetime translations, whose group is indicated with $T^{3,1}$. It can be thus indicated as $T^{3,1} \rtimes \, O(3,1)$ (or $ISO(3,1)$ $\equiv$ $T^{3,1} \rtimes \, SO(3,1)$ for transformations continuously connected with the identity). A generic Poincar\'e transformation can be written as:
\be
x'^\mu=\La^\mu_\nu x^\nu+b^\mu \, .
\ee
Therefore the Poincar\'e group is a ten parameters group.

Let us now consider an infinitesimal transformation, to determine the Lie algebra associated with the Poincar\'e group $ISO(3,1)$. Firstly we take into account spacetime translations. An infinitesimal translation can be written as
\be
T(\de b)=\ide+i \de b^\mu P_\mu \, .
\ee
As known $P_\mu$ is the four momentum operator. An infinitesimal Lorentz transformation can be written as
\be
\La(\de \om)=\ide-\frac{i}{2}\de \om_{\mu \, \nu} J^{\mu \, \nu} \, ,
\ee
where $\de \om^{\mu \, \nu}$ is an antisymmetric matrix (has six independent parameters). We have seen that, considering only the spatial indexes, these transformations coincides with $SO(3)$ elements. An infinitesimal rotation can be written as
\be
R(\de \theta)=\ide-i\delta \vartheta_k J^k \, .
\ee
We are then led to do the following identifications:
\be \label{identif1}
\de \vartheta_k \ = \ \varepsilon_{k \, l \, m} \de \om_{l \, m} \, , \,\,\,\,\, \varepsilon_{lmk} \, J^k \ = \ -J_{l \, m} \, , \,\,\,\, k,l,m=1,2,3 \, .
\ee
In the same way a Lorentz boost can be written as
\be
\La(\de \xi)=\ide-i\delta \xi^k K_k \, ,
\ee
identifying
\be \label{identif2}
\de \xi^m \ = \ \de \om_{0 \, m} \, , \,\,\,\,\, K_m \ = \ J^{0 \, m} \, , \,\,\,\, m=1,2,3 \, .
\ee
One can thus derive the Poincar\'e algebra:
\bea \label{poal1}
&& [P_\mu,P_\la]=0 \, ,\\[2mm] \label{poal2}
&& [P_\mu,J_{\la \si}]=i (P_\la g_{\mu \si}-P_\si g_{\mu \la}) \, , \\[2mm] \label{poal3}
&& [J_{\mu \nu},J_{\la \si}]=i (J_{\la \nu} g_{\mu \si}-J_{\si \nu} g_{\mu \la}+J_{\mu \la} g_{\nu \si}-J_{\mu \si} g_{\nu \la}) \, .
\eea
%
\section{Time evolution of flavor ladder operators}\label{ApA}
We here report the explicit form of the functions $f^\G k_{\si \rho}$ and $g^\G k_{\si \rho}$ introduced in Eq,\eqref{fg}. By using Eq.\eqref{4x4Bog} we get:
\bea
f^\G k_{AA}(t) & = &  \cos^2 \theta \,\frac{\om_{\G k,1}}{\om_{\G k,A}}\, \lf(|\rho^\G k_{A 1}|^2 \, e^{i\lf(\om_{\G k,A}-\om_{\G k,1}\ri)t}- |\la^\G k_{A 1}|^2 e^{i\lf(\om_{\G k,A}+\om_{\G k,1}\ri)t}\ri) \\[2mm] \non
&+& \sin^2 \theta \, \frac{\om_{\G k,2}}{\om_{\G k,A}} \, \lf(|\rho^\G k_{A 2}|^2 \, e^{i\lf(\om_{\G k,A}-\om_{\G k,2}\ri)t}- |\la^\G k_{A 2}|^2 e^{i\lf(\om_{\G k,A}+\om_{\G k,2}\ri)t}\ri) \, ,\\[2mm]
f^\G k_{BB}(t) & = & \sin^2 \theta \, \frac{\om_{\G k,1}}{\om_{\G k,B}}\, \lf(|\rho^\G k_{B 1}|^2 \, e^{i\lf(\om_{\G k,B}-\om_{\G k,1}\ri)t}- |\la^\G k_{B 1}|^2 e^{i\lf(\om_{\G k,B}+\om_{\G k,1}\ri)t}\ri) \\[2mm] \non
&+& \cos^2 \theta \, \frac{\om_{\G k,2}}{\om_{\G k,B}} \, \lf(|\rho^\G k_{B 2}|^2 \, e^{i\lf(\om_{\G k,B}-\om_{\G k,2}\ri)t}- |\la^\G k_{B 2}|^2 e^{i\lf(\om_{\G k,B}+\om_{\G k,2}\ri)t}\ri) \, , \\[2mm]
f^\G k_{\si\rho}(t) & = & \frac{\sin \theta \, \cos \theta}{\om_{\G k,\rho}}  \, \sum^2_{j=1}(-1)^j\,\om_{\G k,j}\,  \lf(|\rho^\G k_{\si j}||\rho^\G k_{\rho j}| \, e^{i\lf(\om_{\G k,\si}-\om_{\G k,j}\ri)t}- |\la^\G k_{\si j}||\la^\G k_{\rho j}| e^{i\lf(\om_{\G k,\si}+\om_{\G k,j}\ri)t}\ri) \, \quad \si \ \neq \ \rho \, , \\[2mm]
g^\G k_{AA}(t) & = & \cos^2 \theta \, \frac{\om_{\G k,1}}{\om_{\G k,A}} \, \lf(|\rho^\G k_{A 1}| |\la^\G k_{A 1}|\, e^{i (\om_{\G k,A}+\om_{\G k,1})t}-|\rho^\G k_{A 1}| |\la^\G k_{A 1}| \, e^{i (\om_{\G k,A}-\om_{\G k,1})t}\ri)\non \\[2mm]
& + & \sin^2 \theta \, \frac{\om_{\G k,2}}{\om_{\G k,A}} \, \lf(|\rho^\G k_{A 2}| |\la^\G k_{A 2}|\, e^{i (\om_{\G k,A}+\om_{\G k,2})t}-|\rho^\G k_{A 2}| |\la^\G k_{A 2}| \, e^{i (\om_{\G k,A}-\om_{\G k,2})t}\ri) \, , \\[2mm]
g^\G k_{BB}(t) & = & \sin^2 \theta \, \frac{\om_{\G k,1}}{\om_{\G k,B}} \, \lf(|\rho^\G k_{B 1}| |\la^\G k_{B 1}|\, e^{i (\om_{\G k,B}+\om_{\G k,1})t}-|\rho^\G k_{B 1}| |\la^\G k_{B 1}| \, e^{i (\om_{\G k,B}-\om_{\G k,1})t}\ri)\non \\[2mm]
& + & \cos^2 \theta \, \frac{\om_{\G k,2}}{\om_{\G k,B}} \, \lf(|\rho^\G k_{B 2}| |\la^\G k_{B 2}|\, e^{i (\om_{\G k,B}+\om_{\G k,2})t}-|\rho^\G k_{B 2}| |\la^\G k_{B 2}| \, e^{i (\om_{\G k,B}-\om_{\G k,2})t}\ri) \, , \\[2mm]
g^\G k_{\si\rho}(t) & = & \frac{\sin \theta \, \cos \theta}{\om_{\G k,\rho}}  \, \sum^2_{j=1}(-1)^j\,\om_{\G k,j}\, \lf(|\rho^\G k_{\si j}||\la^\G k_{\rho j}| \, e^{i\lf(\om_{\G k,\si}-\om_{\G k,j}\ri)t}-|\la^\G k_{\si j}||\rho^\G k_{\rho j}| e^{i\lf(\om_{\G k,\si}+\om_{\G k,j}\ri)t}\ri) \, \quad \si \ \neq \ \rho \, . \\[2mm]
\eea
At $t=0$ we have $f_{\si\rho}(0) \ = \ \de_{\si\rho}$ and $ g_{\si\rho}(0)\ = \ 0$ as we would expect. Moreover, the functions $w^{\G k}_{\si\rho}$ and $y^{\G k}_{\si\rho}$ introduced in Eq.\eqref{finham}, read:
\bea \label{w1}
w^{\G k}_{AA} & = & \frac{\om^2_{\G k,1}+\om^2_{\G k,2}+2 \om^2_{\G k,A}+\cos 2 \theta \lf(\om^2_{\G k,1}-\om^2_{\G k,2}\ri)}{4 \, \om_{\G k,A}} \, ,\\[2mm] \label{w2}
w^{\G k}_{BB} & = & \frac{\om^2_{\G k,1}+\om^2_{\G k,2}+2 \om^2_{\G k,B}-\cos 2 \theta \lf(\om^2_{\G k,1}-\om^2_{\G k,2}\ri)}{4 \, \om_{\G k,B}} \, , \\[2mm] \label{w3}
w^{\G k}_{\si\rho} & = & y^{\G k}_{\rho \si} \ = \  \frac{\sin 2 \theta \lf(\om^2_{\G k,2}-\om^2_{\G k,1}\ri)}{4 \, \om_{\G k,\rho}} \, \quad \si \ \neq \ \rho  \, , \\[2mm] \label{y1}
y^{\G k}_{AA} & = & \frac{\om^2_{\G k,1}+\om^2_{\G k,2}-2 \om^2_{\G k,A}+\cos 2 \theta \lf(\om^2_{\G k,1}-\om^2_{\G k,2}\ri)}{4 \, \om_{\G k,A}} \, , \\[2mm] \label{y2}
y^{\G k}_{BB} & = & \frac{\om^2_{\G k,1}+\om^2_{\G k,2}-2 \om^2_{\G k,B}-\cos 2 \theta \lf(\om^2_{\G k,1}-\om^2_{\G k,2}\ri)}{4 \, \om_{\G k,B}} \, .
\eea
Note that when there is no mixing $ w^{\G k}_{\si \si}=\om_{\G k,\si}$ and the other coefficients go to zero, as expected.
\section*{References}


\begin{thebibliography}{99}

\bibitem{BarWig}
V.~Bargmann, E.P.~Wigner,
Proc.\ Natl. \ Acad. \ Sci. \ U.S.A. {\bf 34}, 211 (1948).

\bibitem{StWi}
R.F.~Streater and A.S.~Wightman,
\textit{PCT, Spin and Statistics and all that}, (W.A.Benjamin, New York, 1964).

\bibitem{BogLog}
N.N.~Bogoliubov, A.A.~Logunov, A.I.~Oksak and I.~Todorov, \textit{General Principles of Quantum Field Theory}, (Kluwer Academic Publishers, Dordrecht, 1990).

\bibitem{Perkins:2000}
D.H.~Perkins, \textit{Introduction to High Energy Physics}, (Cambridge University Press, Cambridge, 2000 )

\bibitem{uns}
K.~Bhattacharyya,
J. Phys. A {\bf 16}, 2993 (1983);
D.J.~Griffiths,
{\it Introduction to Quantum Mechanics}, (Prentice Hall, New Jersey, 1995).

\bibitem{DeFVit}
 S.~De Filippo and G.~Vitiello,
  Lett.\ Nuovo Cim.\  {\bf 19}, 92 (1977).

\bibitem{BJSun}
M.~Blasone, P.~Jizba and L.~Smaldone,
Phys. \ Rev. \ D. {\bf 99}, 016014 (2019).


\bibitem{Bil}	
S.M.~Bilenky,
Phys. Scripta T {\bf 127}, 8 (2006);
S.M~Bilenky and M.D.~Mateev,
Phys. Part. Nucl. {\bf 38}, 117 (2007);
S.M~Bilenky, F.~von Feilitzsch and W.~Potzel,	
J. Phys. G {\bf 35}, 095003 (2008);
S.M~Bilenky, F.~von Feilitzsch and W.~Potzel,	
J. Phys. G {\bf 38}, 115002 (2011).


\bibitem{Blasone:2019jtj}
  M.~Blasone, G.~Lambiase, G.~G.~Luciano, L.~Petruzziello and L.~Smaldone,
  arXiv:1904.05261 [hep-th].
	
	
	
	
	
	\bibitem{Lobanov}
A.E~Lobanov
 Ann. \ Phys. \ {\bf 403}, 82 (2019).

	\bibitem{ColMan}
	S.~Coleman and T.~Mandula,
	Phys. \ Rev. {\bf 159}, 159 (1967).

\bibitem{Giunti:2003ku}
  C.~Giunti,
  Am. \ J. \ Phys. \  {\bf 72}, 699 (2004).
	
\bibitem{BLDiMLa}
M.~Blasone, M.~Di Mauro and G.~Lambiase,
Acta \ Phys. \ Polon. \ B {\bf 36},  3255 (2005).
	
\bibitem{maguejo}
M.~Blasone, J.~Magueijo and P.~Pires-Pacheco,
Europhys.\ Lett. {\bf 70}, 600 (2005);
Braz.\ J. \ Phys. {\bf 35}, 447 (2005).
	
\bibitem{Pontecorvo}
S.M.~Bilenky and B.~Pontecorvo,
Phys. Rep. {\bf 41}, 225 (1978);
S.M.~Bilenky and S.T.~Petcov,
Rev. \ Mod. \ Phys. {\bf 59}, 671 (1987).


\bibitem{Giuntibook}
C.~Giunti and C.W.~Kim,
\textit{Fundamentals of Neutrino Physics and Astrophysics} (Oxford Univ. Press, Oxford, 2007)

\bibitem{NeutPheno}
W.~Grimus and P.~Stockinger
Phys. Rev. D {\bf 54}, 3414 (1996);
A.G.~Cohen, S.L.~Glashow and Z.~Ligeti,
Phys. Lett. B {\bf 678}, 191 (2009);
E.K.~Akhmedov and A.Y.~Smirnov,
Phys. Atom. Nucl. {\bf 72}, 1363 (2009);
I.P.~Volobuev,
Int.\ J.\ Mod.\ Phys.\ A {\bf 33}, 1850075 (2018).

\bibitem{MixingReview}
M.~Beuthe,
Phys. Rep. {\bf 375}, 105 (2003);
D.~Kruppke,
\textit{On Theories of Neutrino Oscillations: A Summary and Characterisation of the Problematic Aspects}
CITATION = INSPIRE-1321484.

\bibitem{BlaVit95}
M.~Blasone and G.~Vitiello,
  Annals Phys.\  {\bf 244}, 283 (1995).

\bibitem{qftmixing}
K.~Fujii, C.~Habe and T.~Yabuki,
Phys.\ Rev.\ D {\bf 59}, 113003 (1999);
Phys. \ Rev. \ D {\bf 64}, 013011 (2001);
K.C.~Hannabuss and D.C.~Latimer,
J. \ Phys. \ A {\bf 33}, 1369 (2000);
J. \ Phys. \ A {\bf 36}, L69 (2003);
C.R.~Ji and Y.~Mishchenko,
Phys.\ Rev. \ D {\bf 65}, 096015 (2002);
Ann. \ Phys.\  {\bf 315}, 488 (2005).



\bibitem{Follin:2015}
B.~Follin, L.~Knox, M.~Millea and Z.~Pan, Phys. Rev.
Lett. 115, 091301 (2015).

\bibitem{Ringwald:2004}
A.~Ringwald and Y.Y.Y.~Wong, JCAP 0412, 005 (2004).


\bibitem{Ellis:2004ay}
  J.R.~Ellis, N.E.~Mavromatos and M.~Westmuckett,
  Phys. \ Rev. \ D {\bf 70}, 044036 (2004).
	
	\bibitem{mavroman}
N.E.~Mavromatos and S.~Sarkar,
New \ J. \ Phys. {\bf 10}, 073009 (2008);
N.E~Mavromatos, S.~Sarkar and W.~Tarantino,
Phys. \ Rev. \ D {\bf 80}, 084046 (2009);
Phys. \ Rev. \ D {\bf 84}, 044050 (2011);
Mod. \ Phys. \ Lett. \ A {\bf 28}, 1350045 (2013).


\bibitem{DynMix}
M.~Blasone, P.~Jizba, G.~Lambiase and N.E.~Mavromatos,
J. \ Phys. \ Conf. \ Ser. {\bf 538}, 012003 (2014);
M.~Blasone, P.~Jizba and L.~Smaldone,
Nuovo \ Cim. \ C {\bf 38}, 201 (2015).

\bibitem{ccpaper}
M.~Blasone, P.~Jizba, N.E.~Mavromatos and L.~Smaldone,
Phys. \ Rev. \ D, {\bf 100} 045027  (2019).

\bibitem{ccpaper3}
M.~Blasone, P.~Jizba, N.E.~Mavromatos and L.~Smaldone,
J. Phys. Conf. Series {\bf 1194}, 012014 (2019).
	

	\bibitem{Colladay:1996iz}
  D.~Colladay and V.A.~Kostelecky,
  Phys. \ Rev. \ D {\bf 55}, 6760 (1997);
	Phys. \ Rev. \ D {\rm D} {\bf 58}, 116002 (1998).
	
\bibitem{cptgreen}
 O.~W.~Greenberg,
  Phys.\ Rev.\ Lett.\  {\bf 89}, 231602 (2002).
	
	\bibitem{Kostelecky2004}
  V.A.~Kostelecky and M.~Mewes,
	Phys. \ Rev. \ D {\bf 69}, 016005 (2004).
	
	
	
	
	\bibitem{ColGla}
	S.~Coleman and S.L.~Glashow,
	Phys. \ Lett. \ B {\bf 405}, 249 (1997);
	Phys. \ Rev. \ D {\bf 59}, 116008 (1999).
	
\bibitem{Katori:2012pe}
  T.~Katori [MiniBooNE Collaboration],
  Mod. \ Phys.\ Lett.\ A {\bf 27}, 1230024 (2012).

\bibitem{Antonelli:2018fbv}
  V.~Antonelli, L.~Miramonti and M.D.C.~Torri,
  Eur. \ Phys. \ J. \ C {\bf 78}, 667 (2018).
	
\bibitem{Lambiase:2017adh}
  G.~Lambiase and F.~Scardigli,
  Phys. \ Rev. \ D {\bf 97}, 075003 (2018).
	
	\bibitem{GUP}
	 A.~Kempf, G.~Mangano and R.B.~Mann,
	Phys. \ Rev. \ D {\bf 52}, 1108 (1995);
	F.~Scardigli and R.~Casadio,
	Eur. \ Phys. \ J. \ C {\bf 75}, 425 (2015);
	F.~Scardigli, M.~Blasone, G.G.~Luciano and R.~Casadio,
  Eur. \ Phys. \ J. \ C {\bf 78}, 728 (2018).
	


\bibitem{Binger:1999nj}
  M.~Binger and C.R.~Ji,
  Phys. \ Rev. \ D {\bf 60}, 056005 (1999).	

\bibitem{bosons}
M.~Blasone, A.~Capolupo, O.~Romei and G.~Vitiello,
Phys. \ Rev. \ D {\bf 63} 125015 (2001).

	\bibitem{Blasone:2002wp}
  M.~Blasone, P.~Pires Pacheco and H.~Wan Chan Tseung,
  Phys.\ Rev.\ D {\bf 67}, 073011 (2003).

%

\bibitem{Tung}
W.K.~Tung, {\it Group Theory in Physics}, (World Scientific, Singapore, 1980).

\bibitem{PesSch}
M.E.~Peskin, D.V.~Schroeder,
\textit{An introduction to quantum field theory}, (Westview, 1995)

\bibitem{Curved}
 N.D.~Birrell and P.C.W~Davies,
 {\it Quantum Fields in Curved Space}, (Cambridge U. Press, Cambridge, 1984);
R.~Haag, H.~Narnhofer and U.~Stein,
Commun.\ Math. \ Phys. {\bf 94}, 219 (1984).

\bibitem{Miransky}
V.A~Miransky,
{\it Dynamical Symmetry Breaking in Quantum Field Theories}, (World Scientific, 1993)
	
\bibitem{Casimiro}	
M.~Blasone, G.G.~Luciano, L.~Petruzziello and L.~Smaldone,
Phys. \ Lett. \ B {\bf 786}, 278 (2018).


\bibitem{Blasone:2011zz}
  M.~Blasone,
 J.\ Phys.\ Conf. \ Ser. {\bf 306},  012037 (2011).

\bibitem{BJV}
M~Blasone, P~Jizba and G.~Vitiello,
{\it Quantum Field Theory and its Macroscopic Manifestations} (World Scientific, London, \& ICP, 2011)

\bibitem{FabPic}
E.~Fabri, L.~E.~Picasso,
Phys.\ Rev. \ Lett. {\bf 16}, 408 (1966).

\bibitem{Giacosa:2018dzm}
  F.~Giacosa,
  Adv.\ High Energy Phys.\  {\bf 2018}, 4672051 (2018).

\bibitem{greiner}
W.~Greiner and J.~Reinhardt,
{\it Field Quantization}, (Springer-Verlag, Berlin-Heidelberg, 1996).

\bibitem{ItzZub}
C.~Itzykson and J.B.~Zuber,
{\it Quantum Field Theory}, (McGraw-Hill, Inc., New York, 1980).

\bibitem{Exener}
P.~Exner,
Phys. \ Rev. \ D {\bf 28}, 2621 (1983).

\bibitem{tomboulis} J.~D.~Bjorken,
  Annals Phys.\  {\bf 24}, 174 (1963).
  P.~Kraus and E.~T.~Tomboulis,
  Phys.\ Rev.\ D {\bf 66}, 045015 (2002).


\bibitem{cosmo}
  M.~Blasone, A.~Capolupo, S.~Capozziello, S.~Carloni and G.~Vitiello,
  Phys.\ Lett.\ A {\bf 323}, 182 (2004).

	
\bibitem{chiral}
  K.~Fukushima, D.~E.~Kharzeev and H.~J.~Warringa,
  Phys.\ Rev.\ D {\bf 78}, 074033 (2008).


	
\end{thebibliography}
\end{document}